\documentclass[aps,prapplied,twocolumn,floatfix]{revtex4-2}

\usepackage{mathtools,amsfonts,amssymb}
\usepackage{fixmath}
\usepackage{bm}
\usepackage{nccmath}
\usepackage{xfrac}
\usepackage{xspace}
\usepackage[extdef]{delimset}
\usepackage{graphicx}
\usepackage{xcolor}
\usepackage[inline,shortlabels]{enumitem}
\usepackage{calc}
\usepackage{soul}
\usepackage{hyperref}
\usepackage[capitalize]{cleveref}
\usepackage{siunitx}
\usepackage[casesstyle]{cases}
\usepackage{booktabs,multirow}
\usepackage{csquotes}
\MakeOuterQuote{"} 

\DeclareMathOperator*{\argmin}{arg\,min}

\DeclareMathOperator{\trace}{trace}
\DeclareMathOperator{\diag}{diag}

\begin{document}

\title{Fast, accurate, and predictive method for atom detection\\in site-resolved images of microtrap arrays}

\date{\today}

\author{Marc Cheneau}
\email{marc.cheneau@institutoptique.fr}
\author{Romaric Journet}
\author{Matthieu Boffety}
\author{François Goudail}
\author{Caroline Kulcsár}
\affiliation{Université Paris-Saclay, Institut d'Optique Graduate School, CNRS, Laboratoire Charles Fabry, 91127, Palaiseau, France}

\author{Pauline Trouvé-Peloux}
\affiliation{DTIS, ONERA, Université Paris-Saclay, 91123, Palaiseau, France}

\begin{abstract}
    We introduce a new method, rooted in estimation theory, to detect individual atoms in site-resolved images of microtrap arrays, such as optical lattices or optical tweezers arrays.
    Using labelled test images, we demonstrate drastic improvement of the detection accuracy compared to the popular method based on Wiener deconvolution when the inter-site distance is comparable to the radius of the point spread function.
    The runtime of our method scales approximately linearly with the number of sites, and remains well below \qty{100}{\ms} for an array of \numproduct{100 x 100} sites on a desktop computer. It is therefore fully compatible with a real-time usage.
    Finally, we propose a rigorous definition for the signal-to-noise ratio of the problem, and show that it can be used as a predictor for the detection error rate.
    Our work opens the prospect for future experiments with increased array sizes, or reduced inter-site distances.
\end{abstract}

\maketitle

\section{Introduction}

Over the past 15 years, site-resolved imaging of atoms or molecules in arrays of optical microtraps, whether optical lattices or optical tweezers arrays, has become a central technique for probing spatial correlations in quantum systems~\cite{Browaeys2020,Henriet2020,Gross2021,Kaufman2021}.
In a typical experimental setting, a one- or two-dimensional array loaded with atoms, with an inter-site distance ranging from a few hundred nanometers to a few micrometers, is imaged onto a digital camera with a high-resolution objective.
After calibration of the coordinates of the sites in the image, and of the profile of the point spread function (PSF) of the optical system, the image is processed to detect the presence or absence of an atom in each site.
The reconstruction of the site occupancies is a classical estimation-detection problem~\cite{Kay93,Kay98}, where the signal to be estimated before detection consists in the `brightness' of each site.

When the inter-site distance (\( a \)) is much larger than the PSF radius (\( r_\textsc{psf} \)), a simple binning approach is often sufficient to accurately estimate the site brightnesses~\cite{Kwon2017,Norcia2018,Cooper2018,Madjarov2021}.
Otherwise, one has to resort to more elaborate estimation methods, which we broadly categorize as local non-linear least squares~\cite{Bakr2009,Parsons2015,Omran2015,Alberti2016}, local iterative~\cite{Sherson2010,Schauss2015}, deconvolution~\cite{Miranda2015,Parsons2016,Cheuk2017,Yamamoto2020,Kwon2022,LaRooij2023,Mongkolkiattichai2023,Buob2024}, local maximum likelihood~\cite{MartinezDorantes2017,Burrell2010}, and supervised~\cite{Picard2019,Verstraten2024} or unsupervised~\cite{Impertro2023} neural networks.
Recently, the authors of ref.~\cite{Winklmann2024} benchmarked several methods with a Cramér--Rao bound using labelled test images simulating an optical tweezer array experiment.
They showed that a non-linear least-squares approach achieves a better accuracy than, for instance, the popular Wiener or Richardson--Lucy deconvolution, but at the cost of a prohibitive computation time.

In this article, we employ a generalized Wiener filter approach---which provides the optimal linear estimate of the signal, given the statistical properties of the signal and the noise~\cite{Kay93}---to the reconstruction of the site occupancies.
Using labelled test images, we benchmark our method with the standard Wiener deconvolution, and show that our method drastically improves the detection accuracy when the sites are not optically resolved, that is \( a \lesssim r_\textsc{psf} \).
This means that some settings which were deemed improper for single-atom detection---because of a short inter-site distance or a low site brightness, for instance---can actually be processed efficiently.

Importantly, our method is applicable in real-time, as its runtime scales approximately linearly with the number of sites, and remains well below \qty{100}{\ms} for an array of \numproduct{100 x 100} sites on a desktop computer.
To help the reader come to grips with the numerical implementation, we provide them with a tutorial example in the form of a Jupyter notebook~\cite{Notebook}.

Last but not least, we propose a rigorous definition of the signal-to-noise ratio for the reconstruction problem, i.e.\ a synthetic indicator of the intrinsic `difficulty' of the estimation-detection problem under given experimental conditions.
Furthermore, we show that this indicator is a good predictor for the detection accuracy, and can therefore be used to rationalize the design of the imaging system for future experiments.
For instance, one could determine the minimum numerical aperture (NA) necessary to achieve a target detection error rate;
since a reduced NA usually comes with an increased field of view, larger arrays could be processed in neutral-atom based platforms for quantum computation or quantum simulation~\cite{Gyger2024,Pichard2024,Chiu2025}.
Another possibility is to minimize the inter-site spacing for a given NA, offering a direct benefit to experiments on collective light scattering in tweezer arrays~\cite{Hofer2024}, or dipolar gases in optical lattices~\cite{Su2023}, which require short inter-site distances to make the dipolar interactions sizable.

\section{Model and methodology}%
\label{sec:model}

To quantitatively evaluate the performance of different reconstruction methods, we generate labelled test images using an idealized model~\footnote{More elaborate models exist to simulate specific experimental contexts~\cite{Winklmann2023}, but their number of parameters make them less practical to use.}.
We define a square array with \( N_\text{s} \) sites in an image with \( N_\text{p} \) pixels, and denote \( a \) the inter-site distance (measured in pixels).
The top left site coincides with the center of a pixel unless a global offset is applied to the site positions.

Each site has the same occupancy probability \( p \).
The brightness of the empty sites is zero, while that of the occupied site follows a normal distribution with mean \( \mu \) and variance \( \sigma^2 \).
The finite variance accounts for site-to-site variations of the scattering rate due to non-uniform trap depths or illumination, for instance.
The atoms are treated as independent scatterers, which is a good approximation as long as \( a \) is much larger than the imaging wavelength divided by \( 2\pi \).

Once the brightness of each site has been defined, we build a noiseless image by summing the PSF of each site, multiplied by the corresponding brightness, and adding a uniform background \( k \).
We have chosen a Gaussian PSF for simplicity~\footnote{We have also tested our method with an Airy disk, and obtained very similar results. One can convert the Gaussian HWHM to the radius of the Airy disk by multiplying the former with a factor \num{2.4}, assuming that both functions have the same HWHM}; it is characterized by its half width at half maximum (HWHM), identical for all sites, and normalized such that the area integral of the PSF is equal to unity.
We truncate the PSF when the distance to the center is larger than 3 times the HWHM\@.

Finally, we model the shot noise by treating each pixel of the noiseless image as the mean value of a Poisson distribution.
We also add a normally distributed noise with zero mean and uniform variance \( r^2 \) to account for the read noise of the camera.
This noise model is well suited for CMOS cameras, assuming that the noise induced by the dark current and the digitization are negligible.

The relationship between the site brightnesses and the pixel values in the test image can be summarized in the following matrix equation:
\begin{equation}
    \label{eq:test_image_model}
    \mathbold{y} \sim \mathcal{P}\brk{\mathbold{Mx + k}} + \mathcal{N}\brk{\mathbf{0}, \mathbold{r}^2} \; .
\end{equation}
Here, and in the following, small bold letters denote column vectors, capital bold letters matrices, and (small or capital) normal letters scalars.
The vector \( \mathbold{y} \) represents the pixel values, \( \mathbold{x} \) the brightnesses, \( \mathbold{k} = k \mathbf{1} \) the background, and \( \mathbold{r}^2 = r^2 \mathbf{1} \) the read noise.
\( \mathbold{M} \) is the measurement matrix, with \( \brk{\mathbold{M}}_{ij} \) the integral over the \( i\text{th} \) pixel of the PSF of the \( j\text{th} \) site.
The normalization of the PSF implies that \( \sum_i \brk{\mathbold{M}}_{ij} = 1 \) for all sites \( j \).
\( \mathcal{P}(\mathbold{u}) \) and \( \mathcal{N}(\mathbold{u, v}) \) denote, respectively, the multivariate Poisson distribution with mean \( \mathbold{u} \), and the multivariate normal distribution with mean \( \mathbold{u} \) and diagonal covariance matrix \( \diag(\mathbold{v}) \).
\( \mathbf{1} \) and \( \mathbf{0} \) denote the column vectors of ones and zeros.

An important parameter is the ratio of the inter-site distance (\( a \)) to the PSF HWHM (\( r_\textsc{psf} \)), which determines if neighboring sites are optically resolved (\( a \gg r_\textsc{psf} \)) or not (\(a \lesssim r_\textsc{psf} \)).
The first case is typically encountered in experiments with optical tweezer arrays, while the second characterizes most optical lattice experiments.
We treat both cases in this article, although our primary focus and motivation is for optical lattices.

\begin{figure}
    \centering
    \includegraphics{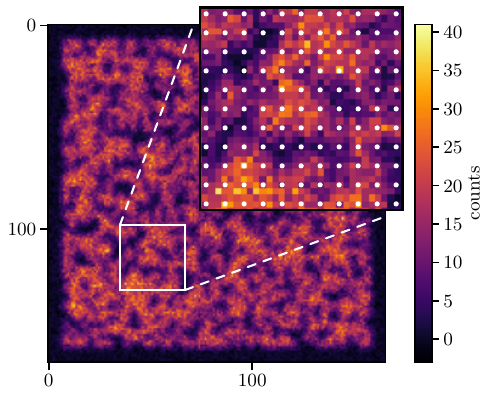}
    \caption{\textbf{Example of test image.}
    The model parameters used to generate the image are given in \cref{tab:params}.
    The inset shows a zoom into the white square, with the white dots indicating the site positions.
    The signal-to-noise ratio of this image is about \qty{15}{\dB}, see \cref{sec:SNR}.
    Our reconstruction method detects the atoms with an error rate of \qty{0.2 +- 0.1}{\percent}, compared to \qty{1.2 +- 0.2}{\percent} with a standard deconvolution estimator, see \cref{sec:deconvolution}.
    The uncertainties represent the standard deviation over \num{1000} images.
    }%
    \label{fig:image}
\end{figure}

\cref{fig:image} shows one typical test image, which we will use throughout the article to illustrate our method.
The parameters used to generate this image are given in \cref{tab:params}.
They were chosen such that the reconstruction is barely possible using conventional techniques.
The corresponding signal-to-noise ratio, as defined in \cref{sec:SNR}, is about \qty{15}{\dB}.
\begin{table}
    \centering
    \begin{ruledtabular}
        \begin{tabular}{l >{\(}c<{\)} >{\(}c<{\)} l}
            number of sites         & N_\text{s}        & \numproduct{50 x 50}  & \\
            occupancy probability   & p                 & 0.6   & \\
            inter-site distance     & a                 & 3     & pixel \\
            PSF HWHM                & r_\text{\sc psf}  & 2     & pixel \\
            brightness mean         & \mu               & \num{200} & count\,/\,site \\
            brightness std.         & \sigma            & \num{20}  & count\,/\,site \\
            background              & k                 & \num{0}   & count\,/\,pixel \\
            read noise std.         & r                 & 1     & count\,/\,pixel \\
        \end{tabular}
    \end{ruledtabular}
    \caption{\textbf{Model parameters for the image in \cref{fig:image}.}
    The unit of length is the pixel size.
    These parameters make a signal-to-noise ratio of about \qty{15}{\dB}, see \cref{sec:SNR}.
    }%
    \label{tab:params}
\end{table}

We quantify the detection accuracy of any reconstruction method using the detection error rate:
\begin{equation}
    \label{eq:DER}
    \text{DER} \triangleq \brk{\text{FP} + \text{FN}} / N_\text{s} \; ,
\end{equation}
where {FP} and {FN} stand for the number of false positives and false negatives, and the terms `positives' and `negatives' refer to sites labelled as occupied or empty, respectively.
The sites are labelled by comparing their estimated brightness to a uniform threshold.
In this work, we have determined the threshold by minimizing the DER\@, knowing the ground truth for each test image~\footnote{In an experimental context, one would rather determine the threshold by first fitting the distribution of estimated brightnesses with a mixture model, and then find the brightness corresponding to an equal likelihood of being empty or filled.}.
For an image like in \cref{fig:image}, our reconstruction method detects the atoms with an error rate of \qty{0.2 +- 0.1}{\percent}, compared to \qty{1.2 +- 0.2}{\percent} with a standard deconvolution estimator, see \cref{sec:deconvolution}.
The uncertainties represent the standard deviation over \num{1000} images.

\section{Optimal linear estimator}%
\label{sec:OLE}

We now turn to the description of our estimation method.
It starts with the choice of a linear model to describe the relationship between the data (\( \mathbold{y} \)) and the variables (\( \mathbold{x} \)):
\begin{equation}
    \label{eq:data_model}
    \mathbold{y = Mx + k + n} \; .
\end{equation}
Here, \( \mathbold{y} \), \( \mathbold{x} \), \( \mathbold{k} \) and \( \mathbold{M} \) have the same meaning as in \cref{eq:test_image_model}.
The vector \( \mathbold{n} \) accounts for the shot noise (originating from both the atomic and background signals) and for the camera read noise;
it has a zero mean value.
\cref{eq:data_model} is relevant in most experimental contexts.
It is even exact for our test images, because the read noise is additive, and it is always possible to split a random vector with a Poisson statistics into its mean value and uncorrelated fluctuations with zero mean value.
In the following, we will always subtract the background from the image, and replace \( \mathbold{y - k} \) with \( \mathbold{y} \) for simplicity.

Ideally, the problem of estimating \( \mathbold{x} \) given \( \mathbold{y} \) should be treated in the maximum likelihood sense with an accurate statistical model for the signal and the noise~\cite{Demoment2008}.
Given the bimodal nature of the brightness probability distribution (empty sites have zero brightness, and occupied sites have a mean brightness \( \mu > 0 \)), the resulting estimation algorithm would be non-linear, therefore time-consuming, and prone to fall into local minima.

To minimize the computation time and ensure the global convergence of the method, we want to limit ourselves to linear estimators.
The optimal linear estimator (OLE), also called generalized Wiener filter, is the matrix \( \mathbold{H} \) which minimizes the mean squared error between the estimated and true values of the variables:
\begin{equation}
    \label{eq:argmin}
    \mathbold{H}_\text{opt} \triangleq \argmin_{\mathbold{H}} \brk[s]{ \text{MSE}\brk{\mathbold{H}} } \; ,
\end{equation}
with
\begin{align}
    \label{eq:MSE}
    \text{MSE}(\mathbold{H}) & \triangleq \avg1{ \norm{ \mathbold{\hat{x}(H) - x} }^2 } \; , \\
    \label{eq:inverse_problem}
    \mathbold{\hat{x}(H)} & \triangleq \mathbold{ H \brk{y - M\avg{x}} + \avg{x}} \; .
\end{align}
Here, \( \norm{\cdot} \) is the Euclidean norm, and \( \avg{\cdot} \) denotes the expectation value over the probability distributions of \( \mathbold{x} \) and \( \mathbold{n} \), which are treated as random variables.

We show in \cref{app:OLE} that the solution to \cref{eq:argmin} can be put in the form
\begin{equation}
    \label{eq:OLE}
    \mathbold{H}_\text{opt} = \mathbold{ \brk{M^\intercal \Sigma_\mathnormal{n}^\mathrm{-1} M + \Sigma_\mathnormal{x}^\mathrm{-1}}^\mathrm{-1} M^\intercal \Sigma_\mathnormal{n}^\mathrm{-1} } \; ,
\end{equation}
where \( \mathbold{\Sigma}_x \triangleq \avg*{ \brk*{\mathbold{x} - \avg{\mathbold{x}}} \brk*{\mathbold{x} - \avg{\mathbold{x}}}^\intercal } \) is the covariance matrix of the variables, and \( \mathbold{\Sigma}_n \triangleq \avg*{ \mathbold{n} \mathbold{n}^\intercal } \) is the covariance matrix of the noise.
To compute \( \hat{\mathbold{x}}\brk{\mathbold{H}_\text{opt}} \) in practice, we solve the linear system
\begin{subequations}%
    \label{eq:linear_system}
    \begin{equation}
        \mathbold{ A \brk{\hat{x} - \avg{x}} = b } \; ,
    \end{equation}
    with
    \begin{align}
        \mathbold{ A } &= \mathbold{ M^\intercal \Sigma_\mathnormal{n}^\mathrm{-1} M + \Sigma_\mathnormal{x}^\mathrm{-1} } \; , \\
        \mathbold{ b } &= \mathbold{ M^\intercal \Sigma_\mathnormal{n}^\mathrm{-1} \brk{y - M\avg{x} } } \; ,
    \end{align}
\end{subequations}
using the conjugate gradient method with an incomplete {LU} preconditioner.
See \cref{app:numerical_implementation} for a brief description of the numerical implementation.
The full code is available online in the form a tutorial example~\cite{Notebook}.

\section{Bayesian interpretation}%
\label{sec:bayes}

The OLE derived in the previous section depends on the mean and covariance matrix of \( \mathbold{x} \) and \( \mathbold{n} \), and therefore on the probability distributions of these quantities.
The more knowledge we have about these distributions, the more accurate will the estimate be.

Let us start by using the model described in \cref{sec:model}, and suppose that the parameters \( p \), \( \mu \) and \( \sigma^2 \) are known. (A method to estimate these parameters using only the recorded images is presented in \cref{sec:learning_prior}.)
The mean and covariance matrices can then be computed using the relations
\begin{subequations}%
    \label{eq:prior}
    \begin{align}
        \label{eq:meanx_prior}
        \avg{\brk{\mathbold{x}}_i}
        &= p \mu \; , \\
        \label{eq:covx_prior}
        \brk{\mathbold{\Sigma}_x}_{ij}
        &= \brk[s]1{ p (1 - p) \mu^2 + p \sigma^2 } \delta_{ij} \; , \\
        \label{eq:covn_prior}
        \brk{\mathbold{\Sigma}_n}_{ij}
        &= \brk[s]1{ r^2 + k + p \mu \brk{\mathbold{M \mathbf{1}}}_i } \delta_{ij} \; ,
    \end{align}
\end{subequations}
where \( \delta_{ij} \) is the Kronecker delta.
Here, each site is assumed to have the same occupancy probability \( p \), which is as much as we can tell before an image is recorded.
We call \( p \) the \emph{prior} probability. 

Once an image is recorded, we gain some information about the probability for each site to be occupied \emph{in this particular sample}: a site which appears bright is more likely to be occupied than a site which appears dark.
Let us assume that we can attribute a \emph{posterior} occupancy probability \( \brk{\mathbold{p}}_i \) to each site \( i \), depending on the apparent site brightness. (We give an explicit protocol to define the posterior occupancy probability in \cref{sec:learning_posterior}).
The mean and covariance matrices can then be computed using the relations
\begin{subequations}%
    \label{eq:posterior}
    \begin{align}
        \label{eq:meanx_posterior}
        \brk{\avg{\mathbold{x}}}_i
        &= \brk{\mathbold{p}}_i \mu \; , \\
        \label{eq:covx_posterior}
        \brk{\mathbold{\Sigma}_x}_{ij}
        &= \brk[c]1{ \brk{\mathbold{p}}_i \brk[s]{1 - \brk{\mathbold{p}}_i} \mu^2 + \brk{\mathbold{p}}_i \sigma^2 } \delta_{ij} \; , \\
        \label{eq:covn_posterior}
        \brk{\mathbold{\Sigma}_n}_{ij}
        &= \brk[s]1{ r^2 + k + \mu \brk{\mathbold{M p}}_i } \delta_{ij} \; .
    \end{align}
\end{subequations}

In the following, we will refer to the estimator obtained from \cref{eq:prior} as being \emph{a priori} optimal, and to the estimator obtained from \cref{eq:posterior} as being \emph{a posteriori} optimal.
We will see that the \emph{a posteriori} OLE is indeed more accurate than the \emph{a priori} OLE\@.

\section{Interpretation in terms of regularized least-squares problem}%
\label{sec:least-squares}

The OLE can also be interpreted as the solution to a weighted and regularized least-squares problem:
\begin{multline}
    \mathbold{\hat{x}} = \argmin_{\mathbold{x}} \big[ \mathbold{ \brk{y - Mx}^\intercal \Sigma_\mathnormal{n}^\mathrm{-1} \brk{y - Mx} } \big. \\[-1ex]
    \big. + \mathbold{ \brk{x - \avg{x}}^\intercal \Sigma_\mathnormal{x}^\mathrm{-1} \brk{x - \avg{x}} } \big] \; .
\end{multline}
The first term of the objective function quantifies the fidelity to the data, weighted by the variance of the noise on each pixel.
The second term regularizes the problem by penalizing the solutions which spread too far away from the expectation \( \avg{\mathbold{x}} \).
It becomes important when the PSFs of neighboring sites strongly overlap, and the standard least-squares problem is ill-posed.
In the regularized least-squares method, one usually treats the regularization operator as a scalar hyperparameter, and optimizes its value based on some criterion (for instance cross-validation).
Here, the Wiener filter approach directly provides us with the optimal operator.

Adopting the point of view of a regularized least-squares problem sheds new light on the difference between the \emph{a priori} and \emph{a posteriori} OLEs.
Imagine that the posterior occupancy probability \( \brk{\mathbold{p}}_i \) of some site \( i\) is close to zero or one.
For this site, the corresponding variance \( \brk{\mathbold{\Sigma}_x}_{ii} \) becomes small, and the regularization forces the estimate \( \brk{\mathbold{x}}_i \) to be close to the expectation \( \brk{\mathbold{p}}_i \mu \).
The result is an effective decoupling of the site \( i \) from its neighbors, which in turn can be estimated more accurately.

\section{Learning about the probability distributions from the data}%
\label{sec:learning}

We now explain how one can determine the probability distributions of \( \mathbold{x} \) and \( \mathbold{n} \) directly from the recorded images.

\subsection{Prior distributions}%
\label{sec:learning_prior}

The first thing to do is estimate \( \avg{x} \triangleq p\mu \) from the sum of all pixel values:
\begin{equation}
    \label{eq:meanx_from_image}
    \avg{x} = \sum_i \brk{\mathbold{y}}_i / N_\text{s} \; .
\end{equation}
We can then compute \( \mathbold{ \avg{y} } = \avg{x} \mathbold{ M } \mathbf{ 1 } \), and remove it from the recorded image.

Next, we replace \( \mathbold{\Sigma}_n \) and \( \mathbold{\Sigma}_x \) by scalar matrices \( \Sigma_n \mathbold{I} \) and \( \Sigma_x \mathbold{I}\) (\( \mathbold{I} \) is the identity matrix).
Within this approximation, the \emph{a priori} OLE takes the simpler form \( \mathbold{H_\text{opt}} = \mathbold{ \brk*{M^\intercal M + \brk{\mathnormal{\Sigma_n / \Sigma_x}} I}^\mathrm{-1} M^\intercal} \).
We treat \( \Sigma_n / \Sigma_x \) as a hyperparameter, denoted \( \gamma \), and tune its value to maximize the separation between empty and occupied sites in the distribution of estimated brightnesses, \( \hat{\mathbold{x}}(\gamma) \).
The rationale behind this strategy comes from the equivalence with a regularized least-square problem put forward in \cref{sec:least-squares}:
if \( \gamma \) is too small, the estimator will tend to amplify the noise, resulting in a smeared distribution;
if \( \gamma \) is too large, the estimator will squeeze the solution around zero.
The optimal value should instead provide the most accurate estimate, and preserve the bimodal character of the brightnesses, leading to a more accurate detection.

To quantify the separation between empty and filled sites, we use the fourth standardized moment (also called kurtosis) of \( \hat{\mathbold{x}}(\gamma) \).
The kurtosis is a convenient alternative to Fisher's linear discriminant for quantifying the separation between the empty and filled sites in the distribution of estimated brightnesses, because it does not require fitting the distribution with a particular model.
We have verified numerically that these two quantities are univocally related once the value of the occupancy probability \( p \) is fixed.

\begin{figure}
    \centering
    \includegraphics{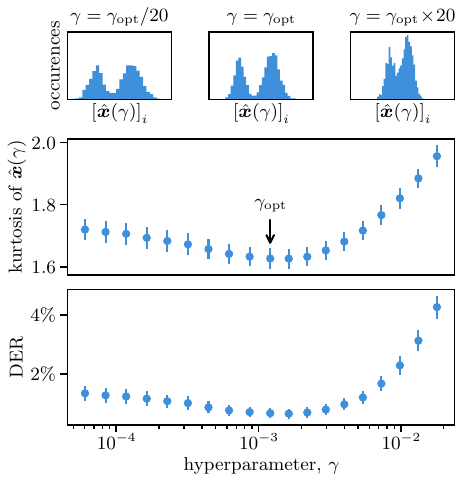}
    \caption{\textbf{Optimizing the hyperparameter of the \emph{a priori} OLE.}
    The optimal value \( \gamma_\text{opt} \) is found by minimizing the  kurtosis of \( \hat{\mathbold{x}}(\gamma) \), which simultaneously minimizes the DER\@.
    Each point is the average over \num{500} test images. The error bars represent the standard deviation. The same set of images is used for all values of \( \gamma \).
    The top panel compares the distributions of \( \hat{\mathbold{x}}(\gamma) \) for a singe test image, and \( \gamma = \gamma_\text{opt} / 20 \), \( \gamma_\text{opt} \), and \( \gamma_\text{opt} \! \times \! 20 \). The three plots are drawn using the same scale for each axis.
    }%
    \label{fig:learning}
\end{figure}

\cref{fig:learning} illustrates the optimization of \( \gamma \), and confirms that the value which maximizes the separation between the empty and filled sites, denoted \( \gamma_\text{opt} \), simultaneously minimizes the DER\@.
This optimal value matches with the expected value \( \Sigma_n / \Sigma_x \), with \( \Sigma_x = \brk{\mathbold{\Sigma}_x}_{ii} \), \(\Sigma_n = \sum_i \brk{\mathbold{\Sigma}_n}_{ii} / N_\text{p} \), and \( \mathbold{\Sigma}_x \), \( \mathbold{\Sigma}_n \) given by \cref{eq:covx_prior,eq:covn_prior}.

Finally, we fit a Gaussian mixture model to \( \hat{\mathbold{x}}(\gamma_\text{opt}) \):
\begin{equation}
    \label{eq:gaussian_mixture}
    \pi\brk{x} = \brk{1 - \phi} \, \mathcal{N}\brk{x; \mu_0, \sigma_0} + \phi \, \mathcal{N}\brk{x; \mu_1, \sigma_1} \; ,
\end{equation}
and determine \( p \), \( \mu \) and \( \sigma^2 \) (see the dashed lines in the top panel of \cref{fig:estimates}).
The first two parameters fall off immediately: \( p \) coincides with the mixture parameter \( \phi \), and \( \mu \) follows from \cref{eq:meanx_from_image}~\footnote{\cref{eq:meanx_from_image} provides a more accurate estimate of \( \mu \) than the Gaussian mixture parameter \( \mu_1 \) because of the `squeezing' effect of the regularization.}.
For \( \sigma^2 \), we can assume that the variance of the Gaussian corresponding to occupied sites (\( \sigma_1^2 \)) is equal to the true variance (\( \sigma^2 \)) plus the estimation errors, which we take equal to the variance of the Gaussian corresponding to empty sites (\( \sigma_0^2 \)).
This reasoning is valid as long as the estimation errors of different sites are not strongly correlated.

\subsection{Posterior distributions}%
\label{sec:learning_posterior}

To determine the posterior distributions, we need to attribute a posterior occupancy probability to each site.
To this end, we exploit the Gaussian mixture model fitted to the \emph{a priori} OLE, and define
\begin{equation}
    \brk{\mathbold{p}}_i \triangleq \frac{\phi \, \mathcal{N}\brk[s]{\brk{\hat{\mathbold{x}}}_i; \mu_1, \sigma_1}}{\brk{1 - \phi} \, \mathcal{N}\brk[s]{\brk{\hat{\mathbold{x}}}_i; \mu_0, \sigma_0} + \phi \, \mathcal{N}\brk[s]{\brk{\hat{\mathbold{x}}}_i; \mu_1, \sigma_1}} \; .
\end{equation}
Then, using the values of \( \mu \) and \( \sigma^2 \) determined earlier, we can compute the first two moments of the posterior distributions using \cref{eq:posterior}.

\begin{figure}
    \centering
    \includegraphics{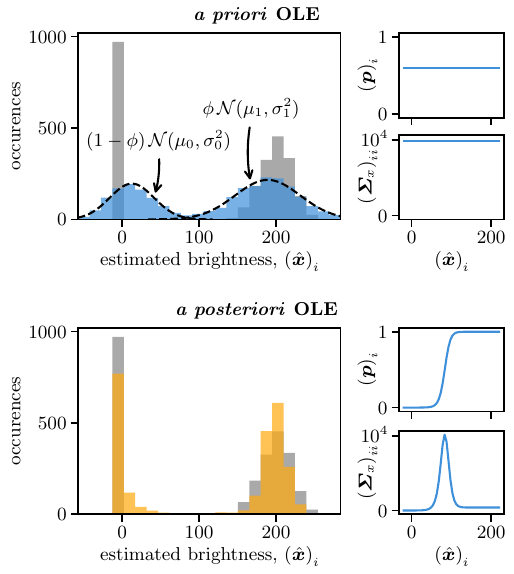}
    \caption{\textbf{OLEs for the test image in \cref{fig:image}.}
    The top and bottom panels show the distribution of estimated brightnesses for the \emph{a priori} and \emph{a posteriori} OLEs, respectively.
    In both panels, the subplots on the right show the underlying occupancy probability and brightness variance, and the gray histogram in the background represents the true brightness distribution.
    In the top panel, the dashed lines are the two modes of the most likely Gaussian mixture distribution, with weights, means and variances \( (1 - \phi), \mu_0, \sigma_0^2 \) (empty sites), and \( \phi, \mu_1, \sigma_1^2 \) (filled sites), see \cref{eq:gaussian_mixture}.
    }%
    \label{fig:estimates}
\end{figure}

In \cref{fig:estimates}, we compare the estimated brightness distributions of the \emph{a priori} and \emph{a posteriori} OLEs for the test image displayed in \cref{fig:image}.
Qualitatively, the distribution obtained with \emph{a posteriori} OLE seems to be closer to the true distribution (shown in gray in the background) than with the \emph{a priori} OLE\@.
Quantitatively, the \emph{a posteriori} OLE reaches \qty{0.2 +- 0.2}{\percent}, compared to \qty{0.7 +- 0.2}{\percent} for the \emph{a priori} OLE, where the uncertainty is the standard deviation over \num{1000} test images generated with the same parameter set.
For comparison, a linear estimator based on the popular Wiener deconvolution filter (see \cref{sec:deconvolution}) gives a DER of \qty{1.2 +- 0.2}{\percent}.

\section{Benchmarking}%
\label{sec:benchmarking}

We now benchmark the \emph{a priori} and \emph{a posteriori} OLEs against a linear estimator based on the popular Wiener deconvolution filter, which we call the `deconvolution estimator'.

\subsection{Deconvolution estimator}%
\label{sec:deconvolution}

The deconvolution estimator proceeds as follows:
The image is first centered around zero by removing the mean pixel value, and filtered using the Wiener deconvolution filter with a uniform (in reciprocal space) regularization operator \( \lambda \in \mathbb{R}^+ \).

To extract the site brightnesses from the filtered image, we perform the convolution by a disk-shaped kernel of radius \( d  \in \mathbb{R}^+ \), and compute the linear interpolation of the result at the site coordinates.
Since the Wiener deconvolution filter, the convolution, and the linear interpolation are all linear operations, the deconvolution estimator as a whole is a linear estimator.

For each set of parameters used in the article, we have performed a joint optimization of the hyperparameters \( \lambda \) and \( d \) by maximizing the kurtosis of \( \hat{\mathbold{x}}(\lambda, d) \), as discussed in \cref{sec:learning_prior}.

\subsection{Detection error rate}%
\label{sec:der}

\begin{figure}
    \centering
    \includegraphics{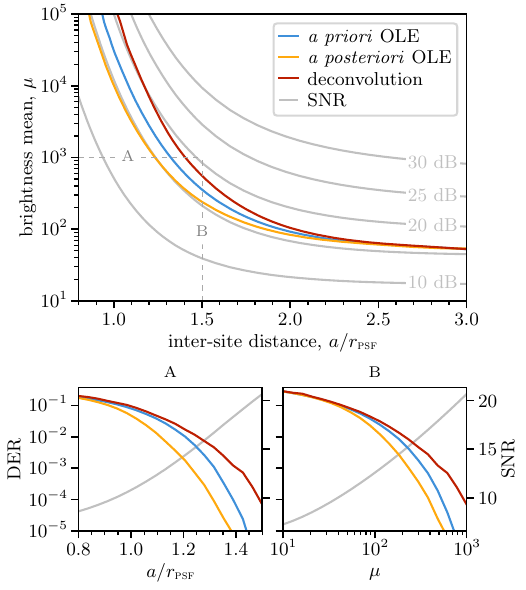}
    \caption{\textbf{Detection error rate and signal-to-noise ratio.}
    Top panel: The colored lines are the DER level curves at \qty{0.1}{\percent} as a function of \( \mu \) and \( a \), with the other parameters as in \cref{tab:params}.
    Each point is averaged over \num{100} test images.
    The gray lines are the SNR level curves at \qtylist{10;15;20;25;30}{\dB}, see \cref{sec:SNR}.
    Lower panel: Cuts through the upper panel along the lines \( \mu = \num{1000} \) (A), and \( a = 1.5 \, r_\textsc{psf} \) (B). The SNR is shown in gray.
    }%
    \label{fig:der}
\end{figure}

\cref{fig:der} shows the level curves at \qty{0.1}{\percent} of each estimator in the space spanned by \( \mu \) and \( a \), with the other parameters fixed to the  values of \cref{tab:params}.
One can clearly distinguish two regimes.
When the sites are optically resolved (\( a \gg r_\textsc{psf} \)), all estimators have exactly the same accuracy.
In this regime, the DER is limited by the fluctuations of the measured signal---due to the variability of the number of scattered photons, or the noise---, rather than by the ability of the estimator to correctly estimates the brightnesses.
In contrast, when the sites are not optically resolved (\( a \lesssim r_\textsc{psf} \)),
the detection accuracies arrange as follows: \( \text{\emph{a posteriori} OLE} > \text{\emph{a priori} OLE} > \text{deconvolution} \), and our method drastically improves the accuracy compared to the standard method.

\subsection{Runtime}

\begin{figure}
    \centering
    \includegraphics{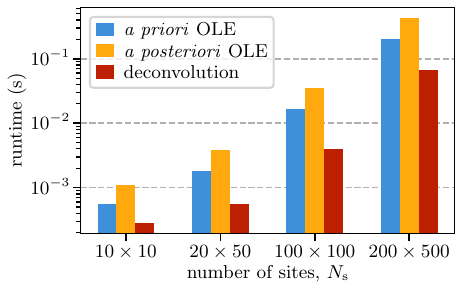}
    \caption{\textbf{Runtime.}
    The runtime is the averaged over \num{5} test images generated with the parameters of \cref{tab:params}, except for the varying number of sites.
    }%
    \label{fig:runtime}
\end{figure}

\cref{fig:runtime} shows the runtime (wall-clock time) of the different estimators for  the parameters of \cref{tab:params} and an increasing number of sites.
While the deconvolution method is always faster, the runtime of the OLEs approximately linearly with the number of sites, and remains well below \qty{100}{\ms} for an array of \numproduct{100 x 100} sites, which makes them both perfectly suitable for real-time usage in all existing experiments.

The runtimes were measured on a desktop computer with an Intel Core {i9-10900K} CPU\@.
Note that the optimization of the hyperparameters and the computation of the Gram matrix are \emph{not} included, because these objects can be cached and reused for a all images recorded under similar experimental conditions.
Note also that the runtime of the \emph{a posteriori} OLE includes that of the \emph{a priori} OLE, since the \emph{a priori} OLE is required to determine the posterior occupancy probability.

\subsection{Robustness against calibration errors}

All estimators used or proposed so far for the reconstruction of the site occupancy assume that the site coordinates and the PSF have been correctly calibrated beforehand. 
In this last section, we study how robust the OLEs and the deconvolution estimator are against two possible calibration errors: an error in the site positions (global offset along one axis of the site array) and an error in the PSF HWHM (global scaling factor)\@.
By calibration error, we mean that the value of the parameter used to compute the estimate is not equal to the value of the parameter which was used to generate the test image.

\begin{figure}
    \centering
    \includegraphics{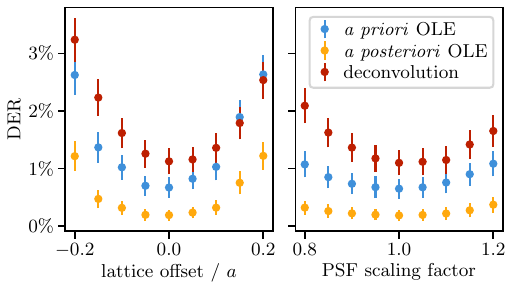}
    \caption{\textbf{Robustness against calibration errors.}
    In the left panel, the coordinates of the sites are affected by a global offset along one axis. The offset is given relative to the inter-site distance \( a \).
    In the right panel, the PSF HWHM is affected by a uniform scaling factor.
    Each point is an average over \num{500} test images generated with the parameters of \cref{tab:params}.
    The error bars represent the standard deviation over the test images.
    }%
    \label{fig:robustness}
\end{figure}

We have computed the DER of each estimator for both types of calibration errors and varying error amplitudes, see \cref{fig:robustness}.
One sees that the OLEs display a comparable or even better robustness against the calibration errors compared to the deconvolution estimator.
They should therefore perform well under realistic experimental conditions.

\section{Signal-to-noise ratio}%
\label{sec:SNR}

A valuable by-product of our approach is a rigorous definition of a signal-to-noise ratio for the estimation-detection problem, namely
\begin{equation}%
    \label{eq:SNR}
    \text{SNR} \triangleq 10 \log_{10} \brk*{ \frac{N_\text{s}\mu^2}{\text{MSE}\brk{\mathbold{H}_\text{opt}}} } \; ,
\end{equation}
with (see \cref{app:OLE})
\begin{equation}
    \label{eq:MSEopt}
    \text{MSE}\brk{\mathbold{H}_\text{opt}} = \trace \brk[s]{ \mathbold{ \brk{ I - H_\text{opt} M } \Sigma_\mathnormal{x} } } \; .
\end{equation}
In the limit \( a \gg r_\textsc{psf} \), the weighted Gram matrix \( \mathbold{M^\intercal \Sigma_\mathnormal{n}^\mathrm{-1} M} \) is diagonal, and no longer depends on \( a / r_\textsc{psf} \).
The optimal MSE then reduces to
\begin{equation}
    \label{eq:SNR_limit}
    \text{MSE}\brk{\mathbold{H}_\text{opt}} = 
    \sum_{i=1}^{N_\text{s}} \brk[s]*{
        1 -
        \frac%
        { \brk{ \mathbold{M^\intercal \Sigma_\mathnormal{n}^\mathrm{-1} M} }_{ii} \brk{ \mathbold{\Sigma}_x^{-1} }_{ii} }%
        { \brk{ \mathbold{M^\intercal \Sigma_\mathnormal{n}^\mathrm{-1} M} }_{ii} + \brk{ \mathbold{\Sigma}_x^{-1} }_{ii} }
    } \; .
\end{equation}

\cref{fig:der} shows level curves of the SNR at \qtylist{10;15;20;25;30}{\dB} (gray lines) alongside the DER level curves.
This graph confirms that the SNR is a good predictor for the DER of both OLEs as their DER level curves are superimposed with SNR level curves in the regimes \( a \gg r_\textsc{psf} \) and \( a \lesssim r_\textsc{psf} \).
In-between these two regimes, a transition occurs.

Being able to predict the DER based on the SNR will be extremely useful to design new experiments as it gives the guarantee of reaching a target detection accuracy without the need to over-design the imaging system.

\section{Conclusion}

We have used the generalized Wiener filter approach to solve the problem of individual atom detection in arrays of optical microtraps.
By design, this method is expected to be the best possible linear estimator for the site brightnesses.
Here, we have shown that it indeed surpasses the standard method based on Wiener deconvolution, drastically reducing the detection error rate when the sites are not optically resolved.

In contrast with maximum likelihood estimators, the computation time remains fully compatible with real-time applications.
It is also robust against calibration errors.
Last but not least, it offers the possibility (not illustrated in this article) to easily account for spatial inhomogeneities in the distribution of atoms, site brightnesses, read noise, or PSF\@.
For instance, one can easily disregard `hot' camera pixels by removing the corresponding lines of the measurement matrix \( \mathbold{M} \), or account for the variations of the PSF across the field of view when the latter is so wide that optical aberrations cannot be neglected.

Benchmarking against supervised or unsupervised machine learning remains to be done.
Artificial neural networks have the ability to precisely model the bimodal brightness probability distribution, but their versatility comes at the cost of a delicate training phase.
On the contrary, our method depends on a few hyperparameters only;
these hyperparameters have a clear physical interpretation, and they can be optimized using only the images to be analyzed.

Finally, the Wiener filter theory provides us with a rigorous definition of a signal-to-noise ratio for the estimation-detection problem.
This SNR aggregates all experimental parameters in a single number, and completely determines the detection error rate.
This opens the prospect for the joint optimization of optical elements and image processing algorithms to improve the global performance of the imaging system in future experiments~\cite{Diaz2009,Burcklen2018}.
Such approach should prove especially useful for experiments aiming at increasing the number of sites or reducing the inter-site distance beyond what is currently feasible.

\begin{acknowledgments}
    The authors thank Félix Faisant, Thomas Chalopin, David Clément, and Isabelle Bouchoule for their feedback on the manuscript, and Anaïs Molineri for early work on the subject.
\end{acknowledgments}

\appendix
\crefalias{section}{appendix}

\section{Derivation of the OLE}%
\label{app:OLE}

In this section we explain how we have derived the expression \labelcref{eq:OLE} of the OLE.
We take \( \avg{\mathbold{x}} = \mathbf{0} \) for simplicity.
We start by writing the squared norm as the trace of the outer product and expand all terms:
\begin{align}
    \text{MSE}(\mathbold{H})
    &= \avg1{ \trace \brk[s]{ \mathbold{\brk{Hy - x}\brk{Hy - x}^\intercal} } } \\
    \label{eq:MSE1}
    &=
    \begin{multlined}[t]
        \trace \bigl[ \mathbold{ H \avg{yy^\intercal} H^\intercal - H \avg{yx^\intercal} }  \bigr. \\ \bigl. \mathbold{ - \avg{xy^\intercal} H^\intercal + \avg{xx^\intercal} } \bigr] \; .
    \end{multlined}
\end{align}
The OLE is such that the derivative of the MSE with respect to \( \mathbold{H} \) is identically zero, which gives~\cite[\namecrefs{eq:MSE}~(100), (104) and (111)]{Petersen2012}:
\begin{equation}
    \eval*{\frac{\mathrm{d} \text{MSE}}{\mathrm{d} \mathbold{H}}}_\mathbold{H_{\mathrlap{\text{opt}}}}
    = 2 \brk[s]{ \mathbold{H_\text{opt} \avg{yy^\intercal} } - \mathbold{\avg{xy^\intercal}} }
    = \bm{0} \; ,
\end{equation}
hence
\begin{equation}\label{eq:MSE2}
    \mathbold{H_\text{opt} = \avg{xy^\intercal}\brk{\avg{yy^\intercal}}^\mathrm{-1} } \; .
\end{equation}
We proceed further by replacing the image vector \( \mathbold{y} \) with \( \mathbold{Mx + n} \):
\begin{align}
    \avg{\mathbold{yx^\intercal}}
    &= \avg{\mathbold{\brk{Mx + n} \, x^\intercal}} \\
    \label{eq:MSE3}
    &= \mathbold{M \avg{xx^\intercal}} \; ,
\end{align}
and
\begin{align}
    \mathbold{ \avg{yy^\intercal} }
    &= \mathbold{ \avg{\brk{Mx + n}\brk{Mx + n}^\intercal} } \\
    \label{eq:MSE4}
    &= \mathbold{ M \avg{xx^\intercal} M^\intercal + \avg{nn^\intercal} } \; ,
\end{align}
In deriving these results, we have used the noise properties \( \mathbold{ \avg{xn^\intercal} = \avg{x} \avg{n^\intercal} } \) (\( \mathbold{x} \) and \( \mathbold{n} \) are uncorrelated), and \( \avg{\mathbold{n}} = \mathbf{0} \).

The matrices \( \avg{\mathbold{xx^\intercal}} \) and \( \avg{\mathbold{nn^\intercal}} \) are the covariance matrices of \( \mathbold{x} \) and \( \mathbold{n} \), denoted \( \mathbold{\Sigma}_x \) and \( \mathbold{\Sigma}_n \) in the main text.
Combining \cref{eq:MSE2,eq:MSE3,eq:MSE4} gives
\begin{align}\label{eq:MSE5}
    \mathbold{ H_\text{opt} = \Sigma_\mathnormal{x} M^\intercal \brk{ M \Sigma_\mathnormal{x} M^\intercal + \Sigma_\mathnormal{n} }^\mathrm{-1} } \; .
\end{align}
This expression turns out to be inconvenient for efficiently computing the solution to the problem.
We obtain the more useful expression in \cref{eq:OLE} by applying the so-called Woodbury identity~\cite[\namecref{eq:MSE}~(158)]{Petersen2012} (which holds because \( \mathbold{\Sigma}_x \) and \( \mathbold{\Sigma}_n \) are both positive definite by definition).

Finally, we can also compute the MSE corresponding to the OLE by combining \cref{eq:MSE1,eq:MSE3,eq:MSE4,eq:MSE5}:
\begin{align}
    \text{MSE}(\mathbold{H_\text{opt}})
    &=
    \begin{multlined}[t]
        \trace \bigl[ \mathbold{ H_\text{opt} \brk{M \Sigma_\mathnormal{x} M^\intercal + \Sigma_\mathnormal{n}} H_\text{opt}^\intercal } \bigr. \\
        \bigl. \mathbold{ - H_\text{opt} M \Sigma_\mathnormal{x} - \Sigma_\mathnormal{x} M^\intercal H_\text{opt}^\intercal + \Sigma_\mathnormal{x} } \bigr]
    \end{multlined} \\
    &= \trace \brk[s]{ \mathbold{ \brk{ I - H_\text{opt} M } \Sigma_\mathnormal{x} } } \; .
\end{align}

\section{Numerical implementation}%
\label{app:numerical_implementation}

We solve the linear system defined by \cref{eq:linear_system} using the preconditioned conjugate gradient (CG) method, which is an iterative algorithm very well suited to large, sparse linear least-square problems.
Here, the sparsity of the matrix \( \mathbold{A} \) follows from the fact that the sites are only coupled to their closest neighbors, and the covariance matrices \( \mathbold{\Sigma}_x \) and \( \mathbold{\Sigma}_n \) are assumed to be diagonal.

The use of a preconditioner is crucial to ensure fast and reliable convergence of the CG algorithm because the condition number of the matrix \( \mathbold{A} \) quickly increases with the ratio \( a / r_\textsc{psf} \).
We have tested different preconditioners (diagonal, incomplete Cholesky and incomplete LU decompositions), and found the Crout version of the incomplete LU ({ILU}) decomposition to be the most efficient and reliable.
We adapt the number of nonzero entries to keep in the LU matrices to the ratio \( a/r_\textsc{psf} \) by checking the convergence of the CG algorithm.
The CG algorithm is stopped when \( \norm{\mathbold{Ax -b}} / \norm{\mathbold{b}} < 10^{-2} \) for all values of \( a/r_\textsc{psf} \).

As the covariance matrix of the noise depends on the occupancy probability (see \cref{eq:covn_prior,eq:covn_posterior}), one should in theory compute the weighted Gram matrix \( \mathbold{ M^\intercal \Sigma_{\mathnormal{n}}^{\mathrm{-1}} M } \) using either the prior or the posterior occupancy probability.
In practice, we found that both the \emph{a priori} and \emph{a posteriori} OLEs obtained with the scalar approximation \( \mathbold{\Sigma}_n = \Sigma_n \mathbold{I} \), \( \Sigma_n = \sum_i \brk{\mathbold{\Sigma}_n}_{ii} / N_\text{p} \), are almost indistinguishable from those obtained without this approximation.
We therefore compute only the non-weighted Gram matrix \( \mathbold{ M^\intercal M } \), and use it for both the \emph{a priori} and \emph{a posteriori} OLEs, which saves computation time.

The numerical implementation is in Python, and relies mainly on the standard libraries NumPy~\cite{Numpy} and SciPy~\cite{Scipy}.
The Wiener deconvolution filter is from the {scikit-image} library~\cite{scikit-image}.
The computation of the incomplete LU decomposition uses the \texttt{ilupp} packages~\cite{ilupp}, which itself is based on the {ILU++} software~\cite{iluplusplus}.
If this package is not available, one may instead rely on the routine provided by Scipy, which itself uses the {SuperLU} library~\cite{superlu}.
We use the Intel Math Kernel Library to speed up sparse matrix-matrix products, with Python bindings provided by the \texttt{sparse\_dot\_mkl} package~\cite{sparse_dot}.

The full code is available online in the form of a tutorial notebook~\cite{Notebook}.

\bibliography{../bibliography.bib}

\begin{thebibliography}{52}%
\makeatletter
\providecommand \@ifxundefined [1]{%
 \@ifx{#1\undefined}
}%
\providecommand \@ifnum [1]{%
 \ifnum #1\expandafter \@firstoftwo
 \else \expandafter \@secondoftwo
 \fi
}%
\providecommand \@ifx [1]{%
 \ifx #1\expandafter \@firstoftwo
 \else \expandafter \@secondoftwo
 \fi
}%
\providecommand \natexlab [1]{#1}%
\providecommand \enquote  [1]{``#1''}%
\providecommand \bibnamefont  [1]{#1}%
\providecommand \bibfnamefont [1]{#1}%
\providecommand \citenamefont [1]{#1}%
\providecommand \href@noop [0]{\@secondoftwo}%
\providecommand \href [0]{\begingroup \@sanitize@url \@href}%
\providecommand \@href[1]{\@@startlink{#1}\@@href}%
\providecommand \@@href[1]{\endgroup#1\@@endlink}%
\providecommand \@sanitize@url [0]{\catcode `\\12\catcode `\$12\catcode
  `\&12\catcode `\#12\catcode `\^12\catcode `\_12\catcode `\%12\relax}%
\providecommand \@@startlink[1]{}%
\providecommand \@@endlink[0]{}%
\providecommand \url  [0]{\begingroup\@sanitize@url \@url }%
\providecommand \@url [1]{\endgroup\@href {#1}{\urlprefix }}%
\providecommand \urlprefix  [0]{URL }%
\providecommand \Eprint [0]{\href }%
\providecommand \doibase [0]{https://doi.org/}%
\providecommand \selectlanguage [0]{\@gobble}%
\providecommand \bibinfo  [0]{\@secondoftwo}%
\providecommand \bibfield  [0]{\@secondoftwo}%
\providecommand \translation [1]{[#1]}%
\providecommand \BibitemOpen [0]{}%
\providecommand \bibitemStop [0]{}%
\providecommand \bibitemNoStop [0]{.\EOS\space}%
\providecommand \EOS [0]{\spacefactor3000\relax}%
\providecommand \BibitemShut  [1]{\csname bibitem#1\endcsname}%
\let\auto@bib@innerbib\@empty
\bibitem [{\citenamefont {Browaeys}\ and\ \citenamefont
  {Lahaye}(2020)}]{Browaeys2020}%
  \BibitemOpen
  \bibfield  {author} {\bibinfo {author} {\bibfnamefont {A.}~\bibnamefont
  {Browaeys}}\ and\ \bibinfo {author} {\bibfnamefont {T.}~\bibnamefont
  {Lahaye}},\ }\bibfield  {title} {\bibinfo {title} {Many-body physics with
  individually controlled rydberg atoms},\ }\href
  {https://doi.org/10.1038/s41567-019-0733-z} {\bibfield  {journal} {\bibinfo
  {journal} {Nature Physics}\ }\textbf {\bibinfo {volume} {16}},\ \bibinfo
  {pages} {132} (\bibinfo {year} {2020})}\BibitemShut {NoStop}%
\bibitem [{\citenamefont {Henriet}\ \emph {et~al.}(2020)\citenamefont
  {Henriet}, \citenamefont {Beguin}, \citenamefont {Signoles}, \citenamefont
  {Lahaye}, \citenamefont {Browaeys}, \citenamefont {Reymond},\ and\
  \citenamefont {Jurczak}}]{Henriet2020}%
  \BibitemOpen
  \bibfield  {author} {\bibinfo {author} {\bibfnamefont {L.}~\bibnamefont
  {Henriet}}, \bibinfo {author} {\bibfnamefont {L.}~\bibnamefont {Beguin}},
  \bibinfo {author} {\bibfnamefont {A.}~\bibnamefont {Signoles}}, \bibinfo
  {author} {\bibfnamefont {T.}~\bibnamefont {Lahaye}}, \bibinfo {author}
  {\bibfnamefont {A.}~\bibnamefont {Browaeys}}, \bibinfo {author}
  {\bibfnamefont {G.-O.}\ \bibnamefont {Reymond}},\ and\ \bibinfo {author}
  {\bibfnamefont {C.}~\bibnamefont {Jurczak}},\ }\bibfield  {title} {\bibinfo
  {title} {Quantum computing with neutral atoms},\ }\href
  {https://doi.org/10.22331/q-2020-09-21-327} {\bibfield  {journal} {\bibinfo
  {journal} {Quantum}\ }\textbf {\bibinfo {volume} {4}},\ \bibinfo {pages}
  {327} (\bibinfo {year} {2020})}\BibitemShut {NoStop}%
\bibitem [{\citenamefont {Gross}\ and\ \citenamefont {Bakr}(2021)}]{Gross2021}%
  \BibitemOpen
  \bibfield  {author} {\bibinfo {author} {\bibfnamefont {C.}~\bibnamefont
  {Gross}}\ and\ \bibinfo {author} {\bibfnamefont {W.~S.}\ \bibnamefont
  {Bakr}},\ }\bibfield  {title} {\bibinfo {title} {Quantum gas microscopy for
  single atom and spin detection},\ }\href
  {https://doi.org/10.1038/s41567-021-01370-5} {\bibfield  {journal} {\bibinfo
  {journal} {Nature Physics}\ }\textbf {\bibinfo {volume} {17}},\ \bibinfo
  {pages} {1316} (\bibinfo {year} {2021})}\BibitemShut {NoStop}%
\bibitem [{\citenamefont {Kaufman}\ and\ \citenamefont
  {Ni}(2021)}]{Kaufman2021}%
  \BibitemOpen
  \bibfield  {author} {\bibinfo {author} {\bibfnamefont {A.~M.}\ \bibnamefont
  {Kaufman}}\ and\ \bibinfo {author} {\bibfnamefont {K.-K.}\ \bibnamefont
  {Ni}},\ }\bibfield  {title} {\bibinfo {title} {Quantum science with optical
  tweezer arrays of ultracold atoms and molecules},\ }\href
  {https://doi.org/10.1038/s41567-021-01357-2} {\bibfield  {journal} {\bibinfo
  {journal} {Nature Physics}\ }\textbf {\bibinfo {volume} {17}},\ \bibinfo
  {pages} {1324} (\bibinfo {year} {2021})}\BibitemShut {NoStop}%
\bibitem [{\citenamefont {Kay}(1993)}]{Kay93}%
  \BibitemOpen
  \bibfield  {author} {\bibinfo {author} {\bibfnamefont {S.~M.}\ \bibnamefont
  {Kay}},\ }\href@noop {} {\emph {\bibinfo {title} {Fundamentals of statistical
  signal processing, Vol. {I}: Estimation Theory}}}\ (\bibinfo  {publisher}
  {Prentice-Hall},\ \bibinfo {year} {1993})\BibitemShut {NoStop}%
\bibitem [{\citenamefont {Kay}(1998)}]{Kay98}%
  \BibitemOpen
  \bibfield  {author} {\bibinfo {author} {\bibfnamefont {S.~M.}\ \bibnamefont
  {Kay}},\ }\href@noop {} {\emph {\bibinfo {title} {Fundamentals of statistical
  signal processing, Vol. {II}: Detection Theory}}}\ (\bibinfo  {publisher}
  {Prentice-Hall},\ \bibinfo {year} {1998})\BibitemShut {NoStop}%
\bibitem [{\citenamefont {Kwon}\ \emph {et~al.}()\citenamefont {Kwon},
  \citenamefont {Ebert}, \citenamefont {Walker},\ and\ \citenamefont
  {Saffman}}]{Kwon2017}%
  \BibitemOpen
  \bibfield  {author} {\bibinfo {author} {\bibfnamefont {M.}~\bibnamefont
  {Kwon}}, \bibinfo {author} {\bibfnamefont {M.~F.}\ \bibnamefont {Ebert}},
  \bibinfo {author} {\bibfnamefont {T.~G.}\ \bibnamefont {Walker}},\ and\
  \bibinfo {author} {\bibfnamefont {M.}~\bibnamefont {Saffman}},\ }\bibfield
  {title} {\bibinfo {title} {Parallel low-loss measurement of multiple atomic
  qubits},\ }\href {https://doi.org/10.1103/PhysRevLett.119.180504} {\bibfield
  {journal} {\bibinfo  {journal} {Physical Review Letters}\ }\textbf {\bibinfo
  {volume} {119}},\ \bibinfo {pages} {180504}}\BibitemShut {NoStop}%
\bibitem [{\citenamefont {Norcia}\ \emph {et~al.}(2018)\citenamefont {Norcia},
  \citenamefont {Young},\ and\ \citenamefont {Kaufman}}]{Norcia2018}%
  \BibitemOpen
  \bibfield  {author} {\bibinfo {author} {\bibfnamefont {M.~A.}\ \bibnamefont
  {Norcia}}, \bibinfo {author} {\bibfnamefont {A.~W.}\ \bibnamefont {Young}},\
  and\ \bibinfo {author} {\bibfnamefont {A.~M.}\ \bibnamefont {Kaufman}},\
  }\bibfield  {title} {\bibinfo {title} {Microscopic {{Control}} and
  {{Detection}} of {{Ultracold Strontium}} in {{Optical}}-{{Tweezer Arrays}}},\
  }\href {https://doi.org/10.1103/PhysRevX.8.041054} {\bibfield  {journal}
  {\bibinfo  {journal} {Physical Review X}\ }\textbf {\bibinfo {volume} {8}},\
  \bibinfo {pages} {041054} (\bibinfo {year} {2018})}\BibitemShut {NoStop}%
\bibitem [{\citenamefont {Cooper}\ \emph {et~al.}(2018)\citenamefont {Cooper},
  \citenamefont {Covey}, \citenamefont {Madjarov}, \citenamefont {Porsev},
  \citenamefont {Safronova},\ and\ \citenamefont {Endres}}]{Cooper2018}%
  \BibitemOpen
  \bibfield  {author} {\bibinfo {author} {\bibfnamefont {A.}~\bibnamefont
  {Cooper}}, \bibinfo {author} {\bibfnamefont {J.~P.}\ \bibnamefont {Covey}},
  \bibinfo {author} {\bibfnamefont {I.~S.}\ \bibnamefont {Madjarov}}, \bibinfo
  {author} {\bibfnamefont {S.~G.}\ \bibnamefont {Porsev}}, \bibinfo {author}
  {\bibfnamefont {M.~S.}\ \bibnamefont {Safronova}},\ and\ \bibinfo {author}
  {\bibfnamefont {M.}~\bibnamefont {Endres}},\ }\bibfield  {title} {\bibinfo
  {title} {Alkaline-{{Earth Atoms}} in {{Optical Tweezers}}},\ }\href
  {https://doi.org/10.1103/PhysRevX.8.041055} {\bibfield  {journal} {\bibinfo
  {journal} {Physical Review X}\ }\textbf {\bibinfo {volume} {8}},\ \bibinfo
  {pages} {041055} (\bibinfo {year} {2018})}\BibitemShut {NoStop}%
\bibitem [{\citenamefont {Madjarov}(2021)}]{Madjarov2021}%
  \BibitemOpen
  \bibfield  {author} {\bibinfo {author} {\bibfnamefont {I.~S.}\ \bibnamefont
  {Madjarov}},\ }\href {https://doi.org/10.7907/d1em-dt34} {\bibinfo {title}
  {Entangling, controlling, and detecting individual strontium atoms in optical
  tweezer arrays}} (\bibinfo {year} {2021})\BibitemShut {NoStop}%
\bibitem [{\citenamefont {Bakr}\ \emph {et~al.}(2009)\citenamefont {Bakr},
  \citenamefont {Gillen}, \citenamefont {Peng}, \citenamefont {Folling},\ and\
  \citenamefont {Greiner}}]{Bakr2009}%
  \BibitemOpen
  \bibfield  {author} {\bibinfo {author} {\bibfnamefont {W.~S.}\ \bibnamefont
  {Bakr}}, \bibinfo {author} {\bibfnamefont {J.~I.}\ \bibnamefont {Gillen}},
  \bibinfo {author} {\bibfnamefont {A.}~\bibnamefont {Peng}}, \bibinfo {author}
  {\bibfnamefont {S.}~\bibnamefont {Folling}},\ and\ \bibinfo {author}
  {\bibfnamefont {M.}~\bibnamefont {Greiner}},\ }\bibfield  {title} {\bibinfo
  {title} {A quantum gas microscope for detecting single atoms in a
  {{Hubbard}}-regime optical lattice},\ }\href
  {https://doi.org/10.1038/nature08482} {\bibfield  {journal} {\bibinfo
  {journal} {Nature}\ }\textbf {\bibinfo {volume} {462}},\ \bibinfo {pages}
  {74} (\bibinfo {year} {2009})}\BibitemShut {NoStop}%
\bibitem [{\citenamefont {Parsons}\ \emph {et~al.}(2015)\citenamefont
  {Parsons}, \citenamefont {Huber}, \citenamefont {Mazurenko}, \citenamefont
  {Chiu}, \citenamefont {Setiawan}, \citenamefont {Wooley-Brown}, \citenamefont
  {Blatt},\ and\ \citenamefont {Greiner}}]{Parsons2015}%
  \BibitemOpen
  \bibfield  {author} {\bibinfo {author} {\bibfnamefont {M.~F.}\ \bibnamefont
  {Parsons}}, \bibinfo {author} {\bibfnamefont {F.}~\bibnamefont {Huber}},
  \bibinfo {author} {\bibfnamefont {A.}~\bibnamefont {Mazurenko}}, \bibinfo
  {author} {\bibfnamefont {C.~S.}\ \bibnamefont {Chiu}}, \bibinfo {author}
  {\bibfnamefont {W.}~\bibnamefont {Setiawan}}, \bibinfo {author}
  {\bibfnamefont {K.}~\bibnamefont {Wooley-Brown}}, \bibinfo {author}
  {\bibfnamefont {S.}~\bibnamefont {Blatt}},\ and\ \bibinfo {author}
  {\bibfnamefont {M.}~\bibnamefont {Greiner}},\ }\bibfield  {title} {\bibinfo
  {title} {Site-{{Resolved Imaging}} of {{Fermionic}} {$^{6}\mathrm{Li}$} in an
  {{Optical Lattice}}},\ }\href
  {https://doi.org/10.1103/PhysRevLett.114.213002} {\bibfield  {journal}
  {\bibinfo  {journal} {Physical Review Letters}\ }\textbf {\bibinfo {volume}
  {114}},\ \bibinfo {pages} {213002} (\bibinfo {year} {2015})}\BibitemShut
  {NoStop}%
\bibitem [{\citenamefont {Omran}\ \emph {et~al.}(2015)\citenamefont {Omran},
  \citenamefont {Boll}, \citenamefont {Hilker}, \citenamefont {Kleinlein},
  \citenamefont {Salomon}, \citenamefont {Bloch},\ and\ \citenamefont
  {Gross}}]{Omran2015}%
  \BibitemOpen
  \bibfield  {author} {\bibinfo {author} {\bibfnamefont {A.}~\bibnamefont
  {Omran}}, \bibinfo {author} {\bibfnamefont {M.}~\bibnamefont {Boll}},
  \bibinfo {author} {\bibfnamefont {T.~A.}\ \bibnamefont {Hilker}}, \bibinfo
  {author} {\bibfnamefont {K.}~\bibnamefont {Kleinlein}}, \bibinfo {author}
  {\bibfnamefont {G.}~\bibnamefont {Salomon}}, \bibinfo {author} {\bibfnamefont
  {I.}~\bibnamefont {Bloch}},\ and\ \bibinfo {author} {\bibfnamefont
  {C.}~\bibnamefont {Gross}},\ }\bibfield  {title} {\bibinfo {title}
  {Microscopic {{Observation}} of {{Pauli Blocking}} in {{Degenerate Fermionic
  Lattice Gases}}},\ }\href {https://doi.org/10.1103/PhysRevLett.115.263001}
  {\bibfield  {journal} {\bibinfo  {journal} {Physical Review Letters}\
  }\textbf {\bibinfo {volume} {115}},\ \bibinfo {pages} {263001} (\bibinfo
  {year} {2015})}\BibitemShut {NoStop}%
\bibitem [{\citenamefont {Alberti}\ \emph {et~al.}(2016)\citenamefont
  {Alberti}, \citenamefont {Robens}, \citenamefont {Alt}, \citenamefont
  {Brakhane}, \citenamefont {Karski}, \citenamefont {Reimann}, \citenamefont
  {{Artur Widera}},\ and\ \citenamefont {Meschede}}]{Alberti2016}%
  \BibitemOpen
  \bibfield  {author} {\bibinfo {author} {\bibfnamefont {A.}~\bibnamefont
  {Alberti}}, \bibinfo {author} {\bibfnamefont {C.}~\bibnamefont {Robens}},
  \bibinfo {author} {\bibfnamefont {W.}~\bibnamefont {Alt}}, \bibinfo {author}
  {\bibfnamefont {S.}~\bibnamefont {Brakhane}}, \bibinfo {author}
  {\bibfnamefont {M.}~\bibnamefont {Karski}}, \bibinfo {author} {\bibfnamefont
  {R.}~\bibnamefont {Reimann}}, \bibinfo {author} {\bibnamefont {{Artur
  Widera}}},\ and\ \bibinfo {author} {\bibfnamefont {D.}~\bibnamefont
  {Meschede}},\ }\bibfield  {title} {\bibinfo {title} {Super-resolution
  microscopy of single atoms in optical lattices},\ }\href
  {https://doi.org/10.1088/1367-2630/18/5/053010} {\bibfield  {journal}
  {\bibinfo  {journal} {New Journal of Physics}\ }\textbf {\bibinfo {volume}
  {18}},\ \bibinfo {pages} {053010} (\bibinfo {year} {2016})}\BibitemShut
  {NoStop}%
\bibitem [{\citenamefont {Sherson}\ \emph {et~al.}(2010)\citenamefont
  {Sherson}, \citenamefont {Weitenberg}, \citenamefont {Endres}, \citenamefont
  {Cheneau}, \citenamefont {Bloch},\ and\ \citenamefont {Kuhr}}]{Sherson2010}%
  \BibitemOpen
  \bibfield  {author} {\bibinfo {author} {\bibfnamefont {J.~F.}\ \bibnamefont
  {Sherson}}, \bibinfo {author} {\bibfnamefont {C.}~\bibnamefont {Weitenberg}},
  \bibinfo {author} {\bibfnamefont {M.}~\bibnamefont {Endres}}, \bibinfo
  {author} {\bibfnamefont {M.}~\bibnamefont {Cheneau}}, \bibinfo {author}
  {\bibfnamefont {I.}~\bibnamefont {Bloch}},\ and\ \bibinfo {author}
  {\bibfnamefont {S.}~\bibnamefont {Kuhr}},\ }\bibfield  {title} {\bibinfo
  {title} {Single-atom resolved fluorescence imaging of an atomic {{Mott}}
  insulator},\ }\href {https://doi.org/10.1038/nature09378} {\bibfield
  {journal} {\bibinfo  {journal} {Nature}\ }\textbf {\bibinfo {volume} {467}},\
  \bibinfo {pages} {68} (\bibinfo {year} {2010})}\BibitemShut {NoStop}%
\bibitem [{\citenamefont {Schauss}(2015)}]{Schauss2015}%
  \BibitemOpen
  \bibfield  {author} {\bibinfo {author} {\bibfnamefont {P.}~\bibnamefont
  {Schauss}},\ }\emph {\bibinfo {title} {High-resolution imaging of ordering in
  Rydberg many-body systems}},\ \href {https://doi.org/10.5282/edoc.18152}
  {Ph.D. thesis},\ \bibinfo  {school} {Ludwig-Maximilians-Universität
  München} (\bibinfo {year} {2015})\BibitemShut {NoStop}%
\bibitem [{\citenamefont {Miranda}\ \emph {et~al.}(2015)\citenamefont
  {Miranda}, \citenamefont {Inoue}, \citenamefont {Okuyama}, \citenamefont
  {Nakamoto},\ and\ \citenamefont {Kozuma}}]{Miranda2015}%
  \BibitemOpen
  \bibfield  {author} {\bibinfo {author} {\bibfnamefont {M.}~\bibnamefont
  {Miranda}}, \bibinfo {author} {\bibfnamefont {R.}~\bibnamefont {Inoue}},
  \bibinfo {author} {\bibfnamefont {Y.}~\bibnamefont {Okuyama}}, \bibinfo
  {author} {\bibfnamefont {A.}~\bibnamefont {Nakamoto}},\ and\ \bibinfo
  {author} {\bibfnamefont {M.}~\bibnamefont {Kozuma}},\ }\bibfield  {title}
  {\bibinfo {title} {Site-resolved imaging of ytterbium atoms in a
  two-dimensional optical lattice},\ }\href
  {https://doi.org/10.1103/PhysRevA.91.063414} {\bibfield  {journal} {\bibinfo
  {journal} {Physical Review A}\ }\textbf {\bibinfo {volume} {91}},\ \bibinfo
  {pages} {063414} (\bibinfo {year} {2015})}\BibitemShut {NoStop}%
\bibitem [{\citenamefont {Parsons}(2016)}]{Parsons2016}%
  \BibitemOpen
  \bibfield  {author} {\bibinfo {author} {\bibfnamefont {M.~F.}\ \bibnamefont
  {Parsons}},\ }\emph {\bibinfo {title} {Probing the Hubbard Model With
  Single-Site Resolution}},\ \href
  {http://nrs.harvard.edu/urn-3:HUL.InstRepos:33493308} {Ph.D. thesis},\
  \bibinfo  {school} {Harvard University} (\bibinfo {year} {2016})\BibitemShut
  {NoStop}%
\bibitem [{\citenamefont {Cheuk}(2017)}]{Cheuk2017}%
  \BibitemOpen
  \bibfield  {author} {\bibinfo {author} {\bibfnamefont {L.~W.}\ \bibnamefont
  {Cheuk}},\ }\emph {\bibinfo {title} {Quantum Gas Microscopy of Strongly
  Correlated Fermions}},\ \href
  {https://quantumgas.mit.edu/wp-content/uploads/2018/12/Cheuk_Lawrence_Thesis.pdf}
  {Ph.D. thesis},\ \bibinfo  {school} {Massachusetts Institute of Technology}
  (\bibinfo {year} {2017})\BibitemShut {NoStop}%
\bibitem [{\citenamefont {Yamamoto}\ \emph {et~al.}(2020)\citenamefont
  {Yamamoto}, \citenamefont {Ozawa}, \citenamefont {Nak}, \citenamefont
  {Nakamura},\ and\ \citenamefont {Fukuhara}}]{Yamamoto2020}%
  \BibitemOpen
  \bibfield  {author} {\bibinfo {author} {\bibfnamefont {R.}~\bibnamefont
  {Yamamoto}}, \bibinfo {author} {\bibfnamefont {H.}~\bibnamefont {Ozawa}},
  \bibinfo {author} {\bibfnamefont {D.~C.}\ \bibnamefont {Nak}}, \bibinfo
  {author} {\bibfnamefont {I.}~\bibnamefont {Nakamura}},\ and\ \bibinfo
  {author} {\bibfnamefont {T.}~\bibnamefont {Fukuhara}},\ }\bibfield  {title}
  {\bibinfo {title} {Single-site-resolved imaging of ultracold atoms in a
  triangular optical lattice},\ }\href
  {https://doi.org/10.1088/1367-2630/abcdc8} {\bibfield  {journal} {\bibinfo
  {journal} {New Journal of Physics}\ }\textbf {\bibinfo {volume} {22}},\
  \bibinfo {pages} {123028} (\bibinfo {year} {2020})}\BibitemShut {NoStop}%
\bibitem [{\citenamefont {Kwon}\ \emph {et~al.}(2022)\citenamefont {Kwon},
  \citenamefont {Kim}, \citenamefont {Hur}, \citenamefont {Huh},\ and\
  \citenamefont {Choi}}]{Kwon2022}%
  \BibitemOpen
  \bibfield  {author} {\bibinfo {author} {\bibfnamefont {K.}~\bibnamefont
  {Kwon}}, \bibinfo {author} {\bibfnamefont {K.}~\bibnamefont {Kim}}, \bibinfo
  {author} {\bibfnamefont {J.}~\bibnamefont {Hur}}, \bibinfo {author}
  {\bibfnamefont {S.}~\bibnamefont {Huh}},\ and\ \bibinfo {author}
  {\bibfnamefont {J.-y.}\ \bibnamefont {Choi}},\ }\bibfield  {title} {\bibinfo
  {title} {Site-resolved imaging of a bosonic {Mott} insulator of
  ${}^7\mathrm{Li}$ atoms},\ }\href
  {https://doi.org/10.1103/PhysRevA.105.033323} {\bibfield  {journal} {\bibinfo
   {journal} {Physical Review A}\ }\textbf {\bibinfo {volume} {105}},\ \bibinfo
  {pages} {033323} (\bibinfo {year} {2022})}\BibitemShut {NoStop}%
\bibitem [{\citenamefont {La~Rooij}\ \emph {et~al.}(2023)\citenamefont
  {La~Rooij}, \citenamefont {Ulm}, \citenamefont {Haller},\ and\ \citenamefont
  {Kuhr}}]{LaRooij2023}%
  \BibitemOpen
  \bibfield  {author} {\bibinfo {author} {\bibfnamefont {A.}~\bibnamefont
  {La~Rooij}}, \bibinfo {author} {\bibfnamefont {C.}~\bibnamefont {Ulm}},
  \bibinfo {author} {\bibfnamefont {E.}~\bibnamefont {Haller}},\ and\ \bibinfo
  {author} {\bibfnamefont {S.}~\bibnamefont {Kuhr}},\ }\bibfield  {title}
  {\bibinfo {title} {A comparative study of deconvolution techniques for
  quantum-gas microscope images},\ }\href
  {https://doi.org/10.1088/1367-2630/aced65} {\bibfield  {journal} {\bibinfo
  {journal} {New Journal of Physics}\ }\textbf {\bibinfo {volume} {25}},\
  \bibinfo {pages} {083036} (\bibinfo {year} {2023})}\BibitemShut {NoStop}%
\bibitem [{\citenamefont {Mongkolkiattichai}\ \emph {et~al.}(2023)\citenamefont
  {Mongkolkiattichai}, \citenamefont {Liu}, \citenamefont {Garwood},
  \citenamefont {Yang},\ and\ \citenamefont {Schauss}}]{Mongkolkiattichai2023}%
  \BibitemOpen
  \bibfield  {author} {\bibinfo {author} {\bibfnamefont {J.}~\bibnamefont
  {Mongkolkiattichai}}, \bibinfo {author} {\bibfnamefont {L.}~\bibnamefont
  {Liu}}, \bibinfo {author} {\bibfnamefont {D.}~\bibnamefont {Garwood}},
  \bibinfo {author} {\bibfnamefont {J.}~\bibnamefont {Yang}},\ and\ \bibinfo
  {author} {\bibfnamefont {P.}~\bibnamefont {Schauss}},\ }\bibfield  {title}
  {\bibinfo {title} {Quantum gas microscopy of fermionic triangular-lattice
  mott insulators},\ }\href {https://doi.org/10.1103/PhysRevA.108.L061301}
  {\bibfield  {journal} {\bibinfo  {journal} {Physical Review A}\ }\textbf
  {\bibinfo {volume} {108}},\ \bibinfo {pages} {L061301} (\bibinfo {year}
  {2023})}\BibitemShut {NoStop}%
\bibitem [{\citenamefont {Buob}\ \emph {et~al.}(2024)\citenamefont {Buob},
  \citenamefont {Höschele}, \citenamefont {Makhalov}, \citenamefont
  {Rubio-Abadal},\ and\ \citenamefont {Tarruell}}]{Buob2024}%
  \BibitemOpen
  \bibfield  {author} {\bibinfo {author} {\bibfnamefont {S.}~\bibnamefont
  {Buob}}, \bibinfo {author} {\bibfnamefont {J.}~\bibnamefont {Höschele}},
  \bibinfo {author} {\bibfnamefont {V.}~\bibnamefont {Makhalov}}, \bibinfo
  {author} {\bibfnamefont {A.}~\bibnamefont {Rubio-Abadal}},\ and\ \bibinfo
  {author} {\bibfnamefont {L.}~\bibnamefont {Tarruell}},\ }\bibfield  {title}
  {\bibinfo {title} {A strontium quantum-gas microscope},\ }\href
  {https://doi.org/10.1103/PRXQuantum.5.020316} {\bibfield  {journal} {\bibinfo
   {journal} {{PRX} Quantum}\ }\textbf {\bibinfo {volume} {5}},\ \bibinfo
  {pages} {020316} (\bibinfo {year} {2024})}\BibitemShut {NoStop}%
\bibitem [{\citenamefont {Martinez-Dorantes}\ \emph {et~al.}(2017)\citenamefont
  {Martinez-Dorantes}, \citenamefont {Alt}, \citenamefont {Gallego},
  \citenamefont {Ghosh}, \citenamefont {Ratschbacher}, \citenamefont
  {Völzke},\ and\ \citenamefont {Meschede}}]{MartinezDorantes2017}%
  \BibitemOpen
  \bibfield  {author} {\bibinfo {author} {\bibfnamefont {M.}~\bibnamefont
  {Martinez-Dorantes}}, \bibinfo {author} {\bibfnamefont {W.}~\bibnamefont
  {Alt}}, \bibinfo {author} {\bibfnamefont {J.}~\bibnamefont {Gallego}},
  \bibinfo {author} {\bibfnamefont {S.}~\bibnamefont {Ghosh}}, \bibinfo
  {author} {\bibfnamefont {L.}~\bibnamefont {Ratschbacher}}, \bibinfo {author}
  {\bibfnamefont {Y.}~\bibnamefont {Völzke}},\ and\ \bibinfo {author}
  {\bibfnamefont {D.}~\bibnamefont {Meschede}},\ }\bibfield  {title} {\bibinfo
  {title} {Fast nondestructive parallel readout of neutral atom registers in
  optical potentials},\ }\href {https://doi.org/10.1103/PhysRevLett.119.180503}
  {\bibfield  {journal} {\bibinfo  {journal} {Physical Review Letters}\
  }\textbf {\bibinfo {volume} {119}},\ \bibinfo {pages} {180503} (\bibinfo
  {year} {2017})}\BibitemShut {NoStop}%
\bibitem [{\citenamefont {Burrell}\ \emph {et~al.}(2010)\citenamefont
  {Burrell}, \citenamefont {Szwer}, \citenamefont {Webster},\ and\
  \citenamefont {Lucas}}]{Burrell2010}%
  \BibitemOpen
  \bibfield  {author} {\bibinfo {author} {\bibfnamefont {A.~H.}\ \bibnamefont
  {Burrell}}, \bibinfo {author} {\bibfnamefont {D.~J.}\ \bibnamefont {Szwer}},
  \bibinfo {author} {\bibfnamefont {S.~C.}\ \bibnamefont {Webster}},\ and\
  \bibinfo {author} {\bibfnamefont {D.~M.}\ \bibnamefont {Lucas}},\ }\bibfield
  {title} {\bibinfo {title} {Scalable simultaneous multiqubit readout with
  $99.99\%$ single-shot fidelity},\ }\href
  {https://doi.org/10.1103/PhysRevA.81.040302} {\bibfield  {journal} {\bibinfo
  {journal} {Physical Review A}\ }\textbf {\bibinfo {volume} {81}},\ \bibinfo
  {pages} {040302} (\bibinfo {year} {2010})}\BibitemShut {NoStop}%
\bibitem [{\citenamefont {Picard}\ \emph {et~al.}(2019)\citenamefont {Picard},
  \citenamefont {Mark}, \citenamefont {Ferlaino},\ and\ \citenamefont
  {Bijnen}}]{Picard2019}%
  \BibitemOpen
  \bibfield  {author} {\bibinfo {author} {\bibfnamefont {L.~R.~B.}\
  \bibnamefont {Picard}}, \bibinfo {author} {\bibfnamefont {M.~J.}\
  \bibnamefont {Mark}}, \bibinfo {author} {\bibfnamefont {F.}~\bibnamefont
  {Ferlaino}},\ and\ \bibinfo {author} {\bibfnamefont {R.~v.}\ \bibnamefont
  {Bijnen}},\ }\bibfield  {title} {\bibinfo {title} {Deep learning-assisted
  classification of site-resolved quantum gas microscope images},\ }\href
  {https://doi.org/10.1088/1361-6501/ab44d8} {\bibfield  {journal} {\bibinfo
  {journal} {Measurement Science and Technology}\ }\textbf {\bibinfo {volume}
  {31}},\ \bibinfo {pages} {025201} (\bibinfo {year} {2019})}\BibitemShut
  {NoStop}%
\bibitem [{\citenamefont {Verstraten}\ \emph {et~al.}(2024)\citenamefont
  {Verstraten}, \citenamefont {Dai}, \citenamefont {Dixmerias}, \citenamefont
  {Peaudecerf}, \citenamefont {Jongh},\ and\ \citenamefont
  {Yefsah}}]{Verstraten2024}%
  \BibitemOpen
  \bibfield  {author} {\bibinfo {author} {\bibfnamefont {J.}~\bibnamefont
  {Verstraten}}, \bibinfo {author} {\bibfnamefont {K.}~\bibnamefont {Dai}},
  \bibinfo {author} {\bibfnamefont {M.}~\bibnamefont {Dixmerias}}, \bibinfo
  {author} {\bibfnamefont {B.}~\bibnamefont {Peaudecerf}}, \bibinfo {author}
  {\bibfnamefont {T.~d.}\ \bibnamefont {Jongh}},\ and\ \bibinfo {author}
  {\bibfnamefont {T.}~\bibnamefont {Yefsah}},\ }\bibfield  {title} {\bibinfo
  {title} {In-situ imaging of a single-atom wave packet in continuous space},\
  }\Eprint {https://arxiv.org/abs/2404.05699} {arXiv:2404.05699 [quant-ph]}
  (\bibinfo {year} {2024})\BibitemShut {NoStop}%
\bibitem [{\citenamefont {Impertro}\ \emph {et~al.}(2023)\citenamefont
  {Impertro}, \citenamefont {Wienand}, \citenamefont {Häfele}, \citenamefont
  {von Raven}, \citenamefont {Hubele}, \citenamefont {Klostermann},
  \citenamefont {Cabrera}, \citenamefont {Bloch},\ and\ \citenamefont
  {Aidelsburger}}]{Impertro2023}%
  \BibitemOpen
  \bibfield  {author} {\bibinfo {author} {\bibfnamefont {A.}~\bibnamefont
  {Impertro}}, \bibinfo {author} {\bibfnamefont {J.~F.}\ \bibnamefont
  {Wienand}}, \bibinfo {author} {\bibfnamefont {S.}~\bibnamefont {Häfele}},
  \bibinfo {author} {\bibfnamefont {H.}~\bibnamefont {von Raven}}, \bibinfo
  {author} {\bibfnamefont {S.}~\bibnamefont {Hubele}}, \bibinfo {author}
  {\bibfnamefont {T.}~\bibnamefont {Klostermann}}, \bibinfo {author}
  {\bibfnamefont {C.~R.}\ \bibnamefont {Cabrera}}, \bibinfo {author}
  {\bibfnamefont {I.}~\bibnamefont {Bloch}},\ and\ \bibinfo {author}
  {\bibfnamefont {M.}~\bibnamefont {Aidelsburger}},\ }\bibfield  {title}
  {\bibinfo {title} {An unsupervised deep learning algorithm for single-site
  reconstruction in quantum gas microscopes},\ }\href
  {https://www.nature.com/articles/s42005-023-01287-w} {\bibfield  {journal}
  {\bibinfo  {journal} {Communications Physics}\ }\textbf {\bibinfo {volume}
  {6}},\ \bibinfo {pages} {1} (\bibinfo {year} {2023})}\BibitemShut {NoStop}%
\bibitem [{\citenamefont {Winklmann}\ \emph {et~al.}(2024)\citenamefont
  {Winklmann}, \citenamefont {Alberti},\ and\ \citenamefont
  {Schulz}}]{Winklmann2024}%
  \BibitemOpen
  \bibfield  {author} {\bibinfo {author} {\bibfnamefont {J.}~\bibnamefont
  {Winklmann}}, \bibinfo {author} {\bibfnamefont {A.}~\bibnamefont {Alberti}},\
  and\ \bibinfo {author} {\bibfnamefont {M.}~\bibnamefont {Schulz}},\
  }\bibfield  {title} {\bibinfo {title} {Comparison of atom detection
  algorithms for neutral atom quantum computing},\ }in\ \href
  {https://doi.org/10.1109/QCE60285.2024.00124} {\emph {\bibinfo {booktitle}
  {2024 {IEEE} International Conference on Quantum Computing and Engineering
  ({QCE})}}},\ Vol.~\bibinfo {volume} {01}\ (\bibinfo {year} {2024})\ pp.\
  \bibinfo {pages} {1048--1057}\BibitemShut {NoStop}%
\bibitem [{Not()}]{Notebook}%
  \BibitemOpen
  \href@noop {} {}\bibinfo {howpublished} {Available at
  \url{https://gitlab.in2p3.fr/gaz-quantiques-lcf/strontium/atom-detection}}\BibitemShut
  {NoStop}%
\bibitem [{\citenamefont {Gyger}\ \emph {et~al.}(2024)\citenamefont {Gyger},
  \citenamefont {Ammenwerth}, \citenamefont {Tao}, \citenamefont {Timme},
  \citenamefont {Snigirev}, \citenamefont {Bloch},\ and\ \citenamefont
  {Zeiher}}]{Gyger2024}%
  \BibitemOpen
  \bibfield  {author} {\bibinfo {author} {\bibfnamefont {F.}~\bibnamefont
  {Gyger}}, \bibinfo {author} {\bibfnamefont {M.}~\bibnamefont {Ammenwerth}},
  \bibinfo {author} {\bibfnamefont {R.}~\bibnamefont {Tao}}, \bibinfo {author}
  {\bibfnamefont {H.}~\bibnamefont {Timme}}, \bibinfo {author} {\bibfnamefont
  {S.}~\bibnamefont {Snigirev}}, \bibinfo {author} {\bibfnamefont
  {I.}~\bibnamefont {Bloch}},\ and\ \bibinfo {author} {\bibfnamefont
  {J.}~\bibnamefont {Zeiher}},\ }\bibfield  {title} {\bibinfo {title}
  {Continuous operation of large-scale atom arrays in optical lattices},\
  }\href {https://doi.org/10.1103/PhysRevResearch.6.033104} {\bibfield
  {journal} {\bibinfo  {journal} {Physical Review Research}\ }\textbf {\bibinfo
  {volume} {6}},\ \bibinfo {pages} {033104} (\bibinfo {year}
  {2024})}\BibitemShut {NoStop}%
\bibitem [{\citenamefont {Pichard}\ \emph {et~al.}(2024)\citenamefont
  {Pichard}, \citenamefont {Lim}, \citenamefont {Bloch}, \citenamefont
  {Vaneecloo}, \citenamefont {Bourachot}, \citenamefont {Both}, \citenamefont
  {M\'{e}riaux}, \citenamefont {Dutartre}, \citenamefont {Hostein},
  \citenamefont {Paris}, \citenamefont {Ximenez}, \citenamefont {Signoles},
  \citenamefont {Browaeys}, \citenamefont {Lahaye},\ and\ \citenamefont
  {Dreon}}]{Pichard2024}%
  \BibitemOpen
  \bibfield  {author} {\bibinfo {author} {\bibfnamefont {G.}~\bibnamefont
  {Pichard}}, \bibinfo {author} {\bibfnamefont {D.}~\bibnamefont {Lim}},
  \bibinfo {author} {\bibfnamefont {E.}~\bibnamefont {Bloch}}, \bibinfo
  {author} {\bibfnamefont {J.}~\bibnamefont {Vaneecloo}}, \bibinfo {author}
  {\bibfnamefont {L.}~\bibnamefont {Bourachot}}, \bibinfo {author}
  {\bibfnamefont {G.-J.}\ \bibnamefont {Both}}, \bibinfo {author}
  {\bibfnamefont {G.}~\bibnamefont {M\'{e}riaux}}, \bibinfo {author}
  {\bibfnamefont {S.}~\bibnamefont {Dutartre}}, \bibinfo {author}
  {\bibfnamefont {R.}~\bibnamefont {Hostein}}, \bibinfo {author} {\bibfnamefont
  {J.}~\bibnamefont {Paris}}, \bibinfo {author} {\bibfnamefont
  {B.}~\bibnamefont {Ximenez}}, \bibinfo {author} {\bibfnamefont
  {A.}~\bibnamefont {Signoles}}, \bibinfo {author} {\bibfnamefont
  {A.}~\bibnamefont {Browaeys}}, \bibinfo {author} {\bibfnamefont
  {T.}~\bibnamefont {Lahaye}},\ and\ \bibinfo {author} {\bibfnamefont
  {D.}~\bibnamefont {Dreon}},\ }\bibfield  {title} {\bibinfo {title}
  {Rearrangement of individual atoms in a 2000-site optical-tweezer array at
  cryogenic temperatures},\ }\href
  {https://doi.org/10.1103/PhysRevApplied.22.024073} {\bibfield  {journal}
  {\bibinfo  {journal} {Physical Review Applied}\ }\textbf {\bibinfo {volume}
  {22}},\ \bibinfo {pages} {024073} (\bibinfo {year} {2024})}\BibitemShut
  {NoStop}%
\bibitem [{\citenamefont {Chiu}\ \emph {et~al.}(2025)\citenamefont {Chiu},
  \citenamefont {Trapp}, \citenamefont {Guo}, \citenamefont {Abobeih},
  \citenamefont {Stewart}, \citenamefont {Hollerith}, \citenamefont
  {Stroganov}, \citenamefont {Kalinowski}, \citenamefont {Geim}, \citenamefont
  {Evered}, \citenamefont {Li}, \citenamefont {Peters}, \citenamefont
  {Bluvstein}, \citenamefont {Wang}, \citenamefont {Greiner}, \citenamefont
  {Vuletić},\ and\ \citenamefont {Lukin}}]{Chiu2025}%
  \BibitemOpen
  \bibfield  {author} {\bibinfo {author} {\bibfnamefont {N.-C.}\ \bibnamefont
  {Chiu}}, \bibinfo {author} {\bibfnamefont {E.~C.}\ \bibnamefont {Trapp}},
  \bibinfo {author} {\bibfnamefont {J.}~\bibnamefont {Guo}}, \bibinfo {author}
  {\bibfnamefont {M.~H.}\ \bibnamefont {Abobeih}}, \bibinfo {author}
  {\bibfnamefont {L.~M.}\ \bibnamefont {Stewart}}, \bibinfo {author}
  {\bibfnamefont {S.}~\bibnamefont {Hollerith}}, \bibinfo {author}
  {\bibfnamefont {P.}~\bibnamefont {Stroganov}}, \bibinfo {author}
  {\bibfnamefont {M.}~\bibnamefont {Kalinowski}}, \bibinfo {author}
  {\bibfnamefont {A.~A.}\ \bibnamefont {Geim}}, \bibinfo {author}
  {\bibfnamefont {S.~J.}\ \bibnamefont {Evered}}, \bibinfo {author}
  {\bibfnamefont {S.~H.}\ \bibnamefont {Li}}, \bibinfo {author} {\bibfnamefont
  {L.~M.}\ \bibnamefont {Peters}}, \bibinfo {author} {\bibfnamefont
  {D.}~\bibnamefont {Bluvstein}}, \bibinfo {author} {\bibfnamefont {T.~T.}\
  \bibnamefont {Wang}}, \bibinfo {author} {\bibfnamefont {M.}~\bibnamefont
  {Greiner}}, \bibinfo {author} {\bibfnamefont {V.}~\bibnamefont {Vuletić}},\
  and\ \bibinfo {author} {\bibfnamefont {M.~D.}\ \bibnamefont {Lukin}},\
  }\bibfield  {title} {\bibinfo {title} {Continuous operation of a coherent
  3,000-qubit system},\ }\Eprint {https://arxiv.org/abs/2506.20660}
  {arXiv:2506.20660 [quant-ph]}  (\bibinfo {year} {2025})\BibitemShut {NoStop}%
\bibitem [{\citenamefont {Hofer}\ \emph {et~al.}(2024)\citenamefont {Hofer},
  \citenamefont {Bloch}, \citenamefont {Biagioni}, \citenamefont {Bonvalet},
  \citenamefont {Browaeys},\ and\ \citenamefont {Ferrier-Barbut}}]{Hofer2024}%
  \BibitemOpen
  \bibfield  {author} {\bibinfo {author} {\bibfnamefont {B.}~\bibnamefont
  {Hofer}}, \bibinfo {author} {\bibfnamefont {D.}~\bibnamefont {Bloch}},
  \bibinfo {author} {\bibfnamefont {G.}~\bibnamefont {Biagioni}}, \bibinfo
  {author} {\bibfnamefont {N.}~\bibnamefont {Bonvalet}}, \bibinfo {author}
  {\bibfnamefont {A.}~\bibnamefont {Browaeys}},\ and\ \bibinfo {author}
  {\bibfnamefont {I.}~\bibnamefont {Ferrier-Barbut}},\ }\bibfield  {title}
  {\bibinfo {title} {Single-atom resolved collective spectroscopy of a
  one-dimensional atomic array},\ }\Eprint {https://arxiv.org/abs/2412.02541}
  {arXiv:2412.02541 [quant-ph]}  (\bibinfo {year} {2024})\BibitemShut {NoStop}%
\bibitem [{\citenamefont {Su}\ \emph {et~al.}(2023)\citenamefont {Su},
  \citenamefont {Douglas}, \citenamefont {Szurek}, \citenamefont {Groth},
  \citenamefont {Ozturk}, \citenamefont {Krahn}, \citenamefont {Hébert},
  \citenamefont {Phelps}, \citenamefont {Ebadi}, \citenamefont {Dickerson},
  \citenamefont {Ferlaino}, \citenamefont {Marković},\ and\ \citenamefont
  {Greiner}}]{Su2023}%
  \BibitemOpen
  \bibfield  {author} {\bibinfo {author} {\bibfnamefont {L.}~\bibnamefont
  {Su}}, \bibinfo {author} {\bibfnamefont {A.}~\bibnamefont {Douglas}},
  \bibinfo {author} {\bibfnamefont {M.}~\bibnamefont {Szurek}}, \bibinfo
  {author} {\bibfnamefont {R.}~\bibnamefont {Groth}}, \bibinfo {author}
  {\bibfnamefont {S.~F.}\ \bibnamefont {Ozturk}}, \bibinfo {author}
  {\bibfnamefont {A.}~\bibnamefont {Krahn}}, \bibinfo {author} {\bibfnamefont
  {A.~H.}\ \bibnamefont {Hébert}}, \bibinfo {author} {\bibfnamefont {G.~A.}\
  \bibnamefont {Phelps}}, \bibinfo {author} {\bibfnamefont {S.}~\bibnamefont
  {Ebadi}}, \bibinfo {author} {\bibfnamefont {S.}~\bibnamefont {Dickerson}},
  \bibinfo {author} {\bibfnamefont {F.}~\bibnamefont {Ferlaino}}, \bibinfo
  {author} {\bibfnamefont {O.}~\bibnamefont {Marković}},\ and\ \bibinfo
  {author} {\bibfnamefont {M.}~\bibnamefont {Greiner}},\ }\bibfield  {title}
  {\bibinfo {title} {Dipolar quantum solids emerging in a hubbard quantum
  simulator},\ }\href {https://doi.org/10.1038/s41586-023-06614-3} {\bibfield
  {journal} {\bibinfo  {journal} {Nature}\ }\textbf {\bibinfo {volume} {622}},\
  \bibinfo {pages} {724} (\bibinfo {year} {2023})}\BibitemShut {NoStop}%
\bibitem [{Note1()}]{Note1}%
  \BibitemOpen
  \bibinfo {note} {More elaborate models exist to simulate specific
  experimental contexts~\cite {Winklmann2023}, but their number of parameters
  make them less practical to use.}\BibitemShut {Stop}%
\bibitem [{Note2()}]{Note2}%
  \BibitemOpen
  \bibinfo {note} {We have also tested our method with an Airy disk, and
  obtained very similar results. One can convert the Gaussian HWHM to the
  radius of the Airy disk by multiplying the former with a factor \num {2.4},
  assuming that both functions have the same HWHM}\BibitemShut {NoStop}%
\bibitem [{Note3()}]{Note3}%
  \BibitemOpen
  \bibinfo {note} {In an experimental context, one would rather determine the
  threshold by first fitting the distribution of estimated brightnesses with a
  mixture model, and then find the brightness corresponding to an equal
  likelihood of being empty or filled.}\BibitemShut {Stop}%
\bibitem [{\citenamefont {Demoment}\ and\ \citenamefont
  {Goussard}(2008)}]{Demoment2008}%
  \BibitemOpen
  \bibfield  {author} {\bibinfo {author} {\bibfnamefont {G.}~\bibnamefont
  {Demoment}}\ and\ \bibinfo {author} {\bibfnamefont {Y.}~\bibnamefont
  {Goussard}},\ }\bibinfo {title} {Inversion within the probabilistic
  framework},\ in\ \href
  {https://doi.org/https://doi.org/10.1002/9780470611197.ch3} {\emph {\bibinfo
  {booktitle} {Bayesian Approach to Inverse Problems}}}\ (\bibinfo  {publisher}
  {John Wiley \& Sons, Ltd},\ \bibinfo {year} {2008})\ Chap.~\bibinfo {chapter}
  {3}, pp.\ \bibinfo {pages} {59--78}\BibitemShut {NoStop}%
\bibitem [{Note4()}]{Note4}%
  \BibitemOpen
  \bibinfo {note} {\protect \cref {eq:meanx_from_image} provides a more
  accurate estimate of \( \mu \) than the Gaussian mixture parameter \( \mu _1
  \) because of the `squeezing' effect of the regularization.}\BibitemShut
  {Stop}%
\bibitem [{\citenamefont {Diaz}\ \emph {et~al.}(2009)\citenamefont {Diaz},
  \citenamefont {Goudail}, \citenamefont {Loiseaux},\ and\ \citenamefont
  {Huignard}}]{Diaz2009}%
  \BibitemOpen
  \bibfield  {author} {\bibinfo {author} {\bibfnamefont {F.}~\bibnamefont
  {Diaz}}, \bibinfo {author} {\bibfnamefont {F.}~\bibnamefont {Goudail}},
  \bibinfo {author} {\bibfnamefont {B.}~\bibnamefont {Loiseaux}},\ and\
  \bibinfo {author} {\bibfnamefont {J.-P.}\ \bibnamefont {Huignard}},\
  }\bibfield  {title} {\bibinfo {title} {Increase in depth of field taking into
  account deconvolution by optimization of pupil mask},\ }\href
  {https://doi.org/10.1364/OL.34.002970} {\bibfield  {journal} {\bibinfo
  {journal} {Opt. Lett.}\ }\textbf {\bibinfo {volume} {34}},\ \bibinfo {pages}
  {2970} (\bibinfo {year} {2009})}\BibitemShut {NoStop}%
\bibitem [{\citenamefont {Burcklen}\ \emph {et~al.}(2018)\citenamefont
  {Burcklen}, \citenamefont {Sauer}, \citenamefont {Diaz},\ and\ \citenamefont
  {Goudail}}]{Burcklen2018}%
  \BibitemOpen
  \bibfield  {author} {\bibinfo {author} {\bibfnamefont {M.-A.}\ \bibnamefont
  {Burcklen}}, \bibinfo {author} {\bibfnamefont {H.}~\bibnamefont {Sauer}},
  \bibinfo {author} {\bibfnamefont {F.}~\bibnamefont {Diaz}},\ and\ \bibinfo
  {author} {\bibfnamefont {F.}~\bibnamefont {Goudail}},\ }\bibfield  {title}
  {\bibinfo {title} {Joint digital-optical design of complex lenses using a
  surrogate image quality criterion adapted to commercial optical design
  software},\ }\href {https://doi.org/10.1364/AO.57.009005} {\bibfield
  {journal} {\bibinfo  {journal} {Applied Optics}\ }\textbf {\bibinfo {volume}
  {57}},\ \bibinfo {pages} {9005} (\bibinfo {year} {2018})}\BibitemShut
  {NoStop}%
\bibitem [{\citenamefont {Petersen}\ and\ \citenamefont
  {Pedersen}(2012)}]{Petersen2012}%
  \BibitemOpen
  \bibfield  {author} {\bibinfo {author} {\bibfnamefont {K.~B.}\ \bibnamefont
  {Petersen}}\ and\ \bibinfo {author} {\bibfnamefont {M.~S.}\ \bibnamefont
  {Pedersen}},\ }\href {http://www2.compute.dtu.dk/pubdb/pubs/3274-full.html}
  {\bibinfo {title} {The matrix cookbook}} (\bibinfo {year} {2012}),\ \bibinfo
  {note} {version 20121115}\BibitemShut {NoStop}%
\bibitem [{\citenamefont {Harris}\ \emph {et~al.}(2020)\citenamefont {Harris},
  \citenamefont {Millman}, \citenamefont {van~der Walt}, \citenamefont
  {Gommers}, \citenamefont {Virtanen}, \citenamefont {Cournapeau},
  \citenamefont {Wieser}, \citenamefont {Taylor}, \citenamefont {Berg},
  \citenamefont {Smith}, \citenamefont {Kern}, \citenamefont {Picus},
  \citenamefont {Hoyer}, \citenamefont {van Kerkwijk}, \citenamefont {Brett},
  \citenamefont {Haldane}, \citenamefont {del R{\'{i}}o}, \citenamefont
  {Wiebe}, \citenamefont {Peterson}, \citenamefont {G{\'{e}}rard-Marchant},
  \citenamefont {Sheppard}, \citenamefont {Reddy}, \citenamefont {Weckesser},
  \citenamefont {Abbasi}, \citenamefont {Gohlke},\ and\ \citenamefont
  {Oliphant}}]{Numpy}%
  \BibitemOpen
  \bibfield  {author} {\bibinfo {author} {\bibfnamefont {C.~R.}\ \bibnamefont
  {Harris}}, \bibinfo {author} {\bibfnamefont {K.~J.}\ \bibnamefont {Millman}},
  \bibinfo {author} {\bibfnamefont {S.~J.}\ \bibnamefont {van~der Walt}},
  \bibinfo {author} {\bibfnamefont {R.}~\bibnamefont {Gommers}}, \bibinfo
  {author} {\bibfnamefont {P.}~\bibnamefont {Virtanen}}, \bibinfo {author}
  {\bibfnamefont {D.}~\bibnamefont {Cournapeau}}, \bibinfo {author}
  {\bibfnamefont {E.}~\bibnamefont {Wieser}}, \bibinfo {author} {\bibfnamefont
  {J.}~\bibnamefont {Taylor}}, \bibinfo {author} {\bibfnamefont
  {S.}~\bibnamefont {Berg}}, \bibinfo {author} {\bibfnamefont {N.~J.}\
  \bibnamefont {Smith}}, \bibinfo {author} {\bibfnamefont {R.}~\bibnamefont
  {Kern}}, \bibinfo {author} {\bibfnamefont {M.}~\bibnamefont {Picus}},
  \bibinfo {author} {\bibfnamefont {S.}~\bibnamefont {Hoyer}}, \bibinfo
  {author} {\bibfnamefont {M.~H.}\ \bibnamefont {van Kerkwijk}}, \bibinfo
  {author} {\bibfnamefont {M.}~\bibnamefont {Brett}}, \bibinfo {author}
  {\bibfnamefont {A.}~\bibnamefont {Haldane}}, \bibinfo {author} {\bibfnamefont
  {J.~F.}\ \bibnamefont {del R{\'{i}}o}}, \bibinfo {author} {\bibfnamefont
  {M.}~\bibnamefont {Wiebe}}, \bibinfo {author} {\bibfnamefont
  {P.}~\bibnamefont {Peterson}}, \bibinfo {author} {\bibfnamefont
  {P.}~\bibnamefont {G{\'{e}}rard-Marchant}}, \bibinfo {author} {\bibfnamefont
  {K.}~\bibnamefont {Sheppard}}, \bibinfo {author} {\bibfnamefont
  {T.}~\bibnamefont {Reddy}}, \bibinfo {author} {\bibfnamefont
  {W.}~\bibnamefont {Weckesser}}, \bibinfo {author} {\bibfnamefont
  {H.}~\bibnamefont {Abbasi}}, \bibinfo {author} {\bibfnamefont
  {C.}~\bibnamefont {Gohlke}},\ and\ \bibinfo {author} {\bibfnamefont {T.~E.}\
  \bibnamefont {Oliphant}},\ }\bibfield  {title} {\bibinfo {title} {Array
  programming with {NumPy}},\ }\href
  {https://doi.org/10.1038/s41586-020-2649-2} {\bibfield  {journal} {\bibinfo
  {journal} {Nature}\ }\textbf {\bibinfo {volume} {585}},\ \bibinfo {pages}
  {357} (\bibinfo {year} {2020})}\BibitemShut {NoStop}%
\bibitem [{\citenamefont {Virtanen}\ \emph {et~al.}(2020)\citenamefont
  {Virtanen}, \citenamefont {Gommers}, \citenamefont {Oliphant}, \citenamefont
  {Haberland}, \citenamefont {Reddy}, \citenamefont {Cournapeau}, \citenamefont
  {Burovski}, \citenamefont {Peterson}, \citenamefont {Weckesser},
  \citenamefont {Bright}, \citenamefont {{van der Walt}}, \citenamefont
  {Brett}, \citenamefont {Wilson}, \citenamefont {Millman}, \citenamefont
  {Mayorov}, \citenamefont {Nelson}, \citenamefont {Jones}, \citenamefont
  {Kern}, \citenamefont {Larson}, \citenamefont {Carey}, \citenamefont {Polat},
  \citenamefont {Feng}, \citenamefont {Moore}, \citenamefont {{VanderPlas}},
  \citenamefont {Laxalde}, \citenamefont {Perktold}, \citenamefont {Cimrman},
  \citenamefont {Henriksen}, \citenamefont {Quintero}, \citenamefont {Harris},
  \citenamefont {Archibald}, \citenamefont {Ribeiro}, \citenamefont
  {Pedregosa}, \citenamefont {{van Mulbregt}},\ and\ \citenamefont {{SciPy 1.0
  Contributors}}}]{Scipy}%
  \BibitemOpen
  \bibfield  {author} {\bibinfo {author} {\bibfnamefont {P.}~\bibnamefont
  {Virtanen}}, \bibinfo {author} {\bibfnamefont {R.}~\bibnamefont {Gommers}},
  \bibinfo {author} {\bibfnamefont {T.~E.}\ \bibnamefont {Oliphant}}, \bibinfo
  {author} {\bibfnamefont {M.}~\bibnamefont {Haberland}}, \bibinfo {author}
  {\bibfnamefont {T.}~\bibnamefont {Reddy}}, \bibinfo {author} {\bibfnamefont
  {D.}~\bibnamefont {Cournapeau}}, \bibinfo {author} {\bibfnamefont
  {E.}~\bibnamefont {Burovski}}, \bibinfo {author} {\bibfnamefont
  {P.}~\bibnamefont {Peterson}}, \bibinfo {author} {\bibfnamefont
  {W.}~\bibnamefont {Weckesser}}, \bibinfo {author} {\bibfnamefont
  {J.}~\bibnamefont {Bright}}, \bibinfo {author} {\bibfnamefont {S.~J.}\
  \bibnamefont {{van der Walt}}}, \bibinfo {author} {\bibfnamefont
  {M.}~\bibnamefont {Brett}}, \bibinfo {author} {\bibfnamefont
  {J.}~\bibnamefont {Wilson}}, \bibinfo {author} {\bibfnamefont {K.~J.}\
  \bibnamefont {Millman}}, \bibinfo {author} {\bibfnamefont {N.}~\bibnamefont
  {Mayorov}}, \bibinfo {author} {\bibfnamefont {A.~R.~J.}\ \bibnamefont
  {Nelson}}, \bibinfo {author} {\bibfnamefont {E.}~\bibnamefont {Jones}},
  \bibinfo {author} {\bibfnamefont {R.}~\bibnamefont {Kern}}, \bibinfo {author}
  {\bibfnamefont {E.}~\bibnamefont {Larson}}, \bibinfo {author} {\bibfnamefont
  {C.~J.}\ \bibnamefont {Carey}}, \bibinfo {author} {\bibfnamefont
  {{\.I}.}~\bibnamefont {Polat}}, \bibinfo {author} {\bibfnamefont
  {Y.}~\bibnamefont {Feng}}, \bibinfo {author} {\bibfnamefont {E.~W.}\
  \bibnamefont {Moore}}, \bibinfo {author} {\bibfnamefont {J.}~\bibnamefont
  {{VanderPlas}}}, \bibinfo {author} {\bibfnamefont {D.}~\bibnamefont
  {Laxalde}}, \bibinfo {author} {\bibfnamefont {J.}~\bibnamefont {Perktold}},
  \bibinfo {author} {\bibfnamefont {R.}~\bibnamefont {Cimrman}}, \bibinfo
  {author} {\bibfnamefont {I.}~\bibnamefont {Henriksen}}, \bibinfo {author}
  {\bibfnamefont {E.~A.}\ \bibnamefont {Quintero}}, \bibinfo {author}
  {\bibfnamefont {C.~R.}\ \bibnamefont {Harris}}, \bibinfo {author}
  {\bibfnamefont {A.~M.}\ \bibnamefont {Archibald}}, \bibinfo {author}
  {\bibfnamefont {A.~H.}\ \bibnamefont {Ribeiro}}, \bibinfo {author}
  {\bibfnamefont {F.}~\bibnamefont {Pedregosa}}, \bibinfo {author}
  {\bibfnamefont {P.}~\bibnamefont {{van Mulbregt}}},\ and\ \bibinfo {author}
  {\bibnamefont {{SciPy 1.0 Contributors}}},\ }\bibfield  {title} {\bibinfo
  {title} {{{SciPy} 1.0: Fundamental Algorithms for Scientific Computing in
  Python}},\ }\href {https://doi.org/10.1038/s41592-019-0686-2} {\bibfield
  {journal} {\bibinfo  {journal} {Nature Methods}\ }\textbf {\bibinfo {volume}
  {17}},\ \bibinfo {pages} {261} (\bibinfo {year} {2020})}\BibitemShut
  {NoStop}%
\bibitem [{\citenamefont {van~der Walt}\ \emph {et~al.}(4 06)\citenamefont
  {van~der Walt}, \citenamefont {{S}ch\"onberger}, \citenamefont
  {{Nunez-Iglesias}}, \citenamefont {{B}oulogne}, \citenamefont {{W}arner},
  \citenamefont {{Y}ager}, \citenamefont {{G}ouillart}, \citenamefont {{Y}u},\
  and\ \citenamefont {the scikit-image contributors}}]{scikit-image}%
  \BibitemOpen
  \bibfield  {author} {\bibinfo {author} {\bibfnamefont {S.}~\bibnamefont
  {van~der Walt}}, \bibinfo {author} {\bibfnamefont {J.~L.}\ \bibnamefont
  {{S}ch\"onberger}}, \bibinfo {author} {\bibfnamefont {J.}~\bibnamefont
  {{Nunez-Iglesias}}}, \bibinfo {author} {\bibfnamefont {F.}~\bibnamefont
  {{B}oulogne}}, \bibinfo {author} {\bibfnamefont {J.~D.}\ \bibnamefont
  {{W}arner}}, \bibinfo {author} {\bibfnamefont {N.}~\bibnamefont {{Y}ager}},
  \bibinfo {author} {\bibfnamefont {E.}~\bibnamefont {{G}ouillart}}, \bibinfo
  {author} {\bibfnamefont {T.}~\bibnamefont {{Y}u}},\ and\ \bibinfo {author}
  {\bibnamefont {the scikit-image contributors}},\ }\bibfield  {title}
  {\bibinfo {title} {scikit-image: image processing in {P}ython},\ }\href
  {https://doi.org/10.7717/peerj.453} {\bibfield  {journal} {\bibinfo
  {journal} {PeerJ}\ }\textbf {\bibinfo {volume} {2}},\ \bibinfo {pages} {e453}
  (\bibinfo {year} {2014-06})}\BibitemShut {NoStop}%
\bibitem [{\citenamefont {Hofreither}()}]{ilupp}%
  \BibitemOpen
  \bibfield  {author} {\bibinfo {author} {\bibfnamefont {C.}~\bibnamefont
  {Hofreither}},\ }\href@noop {} {\bibinfo {title} {ilupp -- {ILU} algorithms
  for {C++} and {Python}}},\ \bibinfo {howpublished}
  {\url{https://github.com/c-f-h/ilupp}}\BibitemShut {NoStop}%
\bibitem [{\citenamefont {Mayer}(2007)}]{iluplusplus}%
  \BibitemOpen
  \bibfield  {author} {\bibinfo {author} {\bibfnamefont {J.}~\bibnamefont
  {Mayer}},\ }\bibfield  {title} {\bibinfo {title} {{ILU}++: A new software
  package for solving sparse linear systems with iterative methods},\ }\href
  {https://doi.org/10.1002/pamm.200700911} {\bibfield  {journal} {\bibinfo
  {journal} {{PAMM}}\ }\textbf {\bibinfo {volume} {7}},\ \bibinfo {pages}
  {2020123} (\bibinfo {year} {2007})}\BibitemShut {NoStop}%
\bibitem [{\citenamefont {Li}(2005)}]{superlu}%
  \BibitemOpen
  \bibfield  {author} {\bibinfo {author} {\bibfnamefont {X.~S.}\ \bibnamefont
  {Li}},\ }\bibfield  {title} {\bibinfo {title} {An overview of {SuperLU}:
  Algorithms, implementation, and user interface},\ }\href
  {https://doi.org/10.1145/1089014.1089017} {\bibfield  {journal} {\bibinfo
  {journal} {{ACM} Trans. Math. Softw.}\ }\textbf {\bibinfo {volume} {31}},\
  \bibinfo {pages} {302} (\bibinfo {year} {2005})}\BibitemShut {NoStop}%
\bibitem [{\citenamefont {Jackson}()}]{sparse_dot}%
  \BibitemOpen
  \bibfield  {author} {\bibinfo {author} {\bibfnamefont {C.~A.}\ \bibnamefont
  {Jackson}},\ }\href@noop {} {\bibinfo {title} {Python wrapper for {Intel Math
  Kernel Library (MKL)} matrix multiplication}},\ \bibinfo {howpublished}
  {\url{https://github.com/flatironinstitute/sparse_dot}}\BibitemShut {NoStop}%
\bibitem [{\citenamefont {Winklmann}\ \emph {et~al.}(2023)\citenamefont
  {Winklmann}, \citenamefont {Tsevas},\ and\ \citenamefont
  {Schulz}}]{Winklmann2023}%
  \BibitemOpen
  \bibfield  {author} {\bibinfo {author} {\bibfnamefont {J.}~\bibnamefont
  {Winklmann}}, \bibinfo {author} {\bibfnamefont {D.}~\bibnamefont {Tsevas}},\
  and\ \bibinfo {author} {\bibfnamefont {M.}~\bibnamefont {Schulz}},\
  }\bibfield  {title} {\bibinfo {title} {Realistic neutral atom image
  simulation},\ }in\ \href {https://doi.org/10.1109/QCE57702.2023.00153} {\emph
  {\bibinfo {booktitle} {2023 {IEEE} International Conference on Quantum
  Computing and Engineering ({QCE})}}},\ Vol.~\bibinfo {volume} {01}\ (\bibinfo
  {year} {2023})\ pp.\ \bibinfo {pages} {1349--1359}\BibitemShut {NoStop}%
\end{thebibliography}%

\end{document}